\RequirePackage{fix-cm}
\documentclass[natbib,smallextended]{svjour3}       
\smartqed  
\usepackage{graphicx}

\usepackage{aps-bibstyle}  
\newcommand{\msun}{\mbox{M$_{\odot}$}}
\newcommand{\Msun}{\mbox{M$_{\odot}$ }}
\newcommand{\lsim}{\raisebox{-0.13cm}{~\shortstack{$<$ \\[-0.07cm]
      $\sim$}}~}
\newcommand{\gsim}{\raisebox{-0.13cm}{~\shortstack{$>$ \\[-0.07cm]
      $\sim$}}~}

\newcounter{species}

\def\lion#1#2#3{\setcounter{species}{#2}#1$\;${\sc\roman{species}$\;\lambda${#3}}\relax}

\def\fion#1#2{[{\setcounter{species}{#2}#1$\;${\sc\roman{species}}\relax}]}
\def\flion#1#2#3{[{\setcounter{species}{#2}#1$\;${\sc\roman{species}}]$\;\lambda${#3}}\relax}
\def\fllion#1#2#3{[{\setcounter{species}{#2}#1$\;${\sc\roman{species}}]$\;\lambda\lambda${#3}}\relax}

 \journalname{Space Science Reviews}
\usepackage{ref_macros} 

\usepackage[usenames,dvipsnames]{color}

\def\Mbh{\ifmmode{M_{\rm \bullet}}\else{$M_{\rm \bullet}$}\fi}
\def\Mwd{\ifmmode{M_{\rm wd}}\else{$M_{\rm wd}$}\fi}
\def\Rwd{\ifmmode{R_{\rm wd}}\else{$R_{\rm wd}$}\fi}
\def\MCh{\ifmmode{M_{\rm Ch}}\else{$M_{\rm Ch}$}\fi}
\def\Rt{\ifmmode{R_{\rm t}}\else{$R_{\rm t}$}\fi}
\def\Rg{\ifmmode{R_{\rm g}}\else{$R_{\rm g}$}\fi}
\def\Rg{\ifmmode{R_{\rm g}}\else{$R_{\rm g}$}\fi}
\def\Rp{\ifmmode{R_{\rm p}}\else{$R_{\rm p}$}\fi}
\def\lsim{\lower0.3em\hbox{$\,\buildrel <\over\sim\,$}}
\def\gsim{\lower0.3em\hbox{$\,\buildrel >\over\sim\,$}}

\begin{document}

\title{Tidal Disruptions of White Dwarfs: Theoretical Models and Observational Prospects\footnote{Published chapter in the book, `The Tidal Disruption of Stars by Massive Black Holes', editors: P.~G.~Jonker, S.~Phinney, E.~M.~Rossi, S.~v.~Velzen, I.~Arcavi \& M.~Falanga}}

\titlerunning{Tidal Disruptions of White Dwarfs}        

\author{Kate Maguire \and
        Michael Eracleous \and
        Peter G.~Jonker \and
        Morgan MacLeod \and
        Stephan Rosswog
}

\authorrunning{Maguire et al.} 

\institute{
            K. Maguire \at 
                School of Physics, 
                Trinity College Dublin, the University of Dublin, College Green, Dublin 2, Ireland.
                \email{kate.maguire@tcd.ie}
                \and
           M. Eracleous \at
              Department of Astronomy \& Astrophysics and Institute for Gravitation and the Cosmos, The Pennsylvania State University, 525 Davey Lab, University Park, PA 16802, USA. 
              \email{mxe17@psu.edu}
              \and
              P.~G.~Jonker \at
              SRON, Netherlands Institute for Space Research, Sorbonnelaan 2, 3584~CA, Utrecht, The Netherlands.\\
              Department of Astrophysics/IMAPP, Radboud University Nijmegen, P.O.~Box 9010, 6500 GL, Nijmegen, The Netherlands.
              \email{p.jonker@astro.ru.nl}
              \and
              M.~MacLeod \at
              Harvard-Smithsonian Center for Astrophysics
              60 Garden Street, MS-51
              Cambridge, MA 02138, USA.
              \email{morgan.macleod@cfa.harvard.edu}
               \and
              S.~Rosswog \at
              Department of Astronomy \& Oskar Klein Centre,
              Stockholm University, AlbaNova, Roslagstullbacken 21, SE-10691 Stockholm, Sweden.
              \email{stephan.rosswog@astro.su.se}
}

\date{Received: 17 May 2019 / Accepted: 15 March 2020}

\maketitle

\begin{abstract}
White dwarf stars that enter the tidal radius of black holes
with masses $\lsim 10^5$ \Msun are doomed to be ripped apart by tidal forces.
 Black holes in this mass range between stellar black holes and supermassive black holes have not been conclusively identified so the detection of a tidal disruption of a white dwarf would provide clear evidence for the existence of intermediate-mass black holes. In this review, we present a theoretical and observational overview of the transient events that result from the tidal disruptions of white dwarfs by intermediate-mass black holes. This includes discussion of the latest simulations and predicted properties, the results of observational searches, as well as a summary of the potential for gravitational wave emission to be detected with upcoming missions.

\end{abstract}

    

\section{Introduction}
\label{sec_intro}

\begin{figure}[t]
    \centerline{\includegraphics[width=0.7\textwidth]{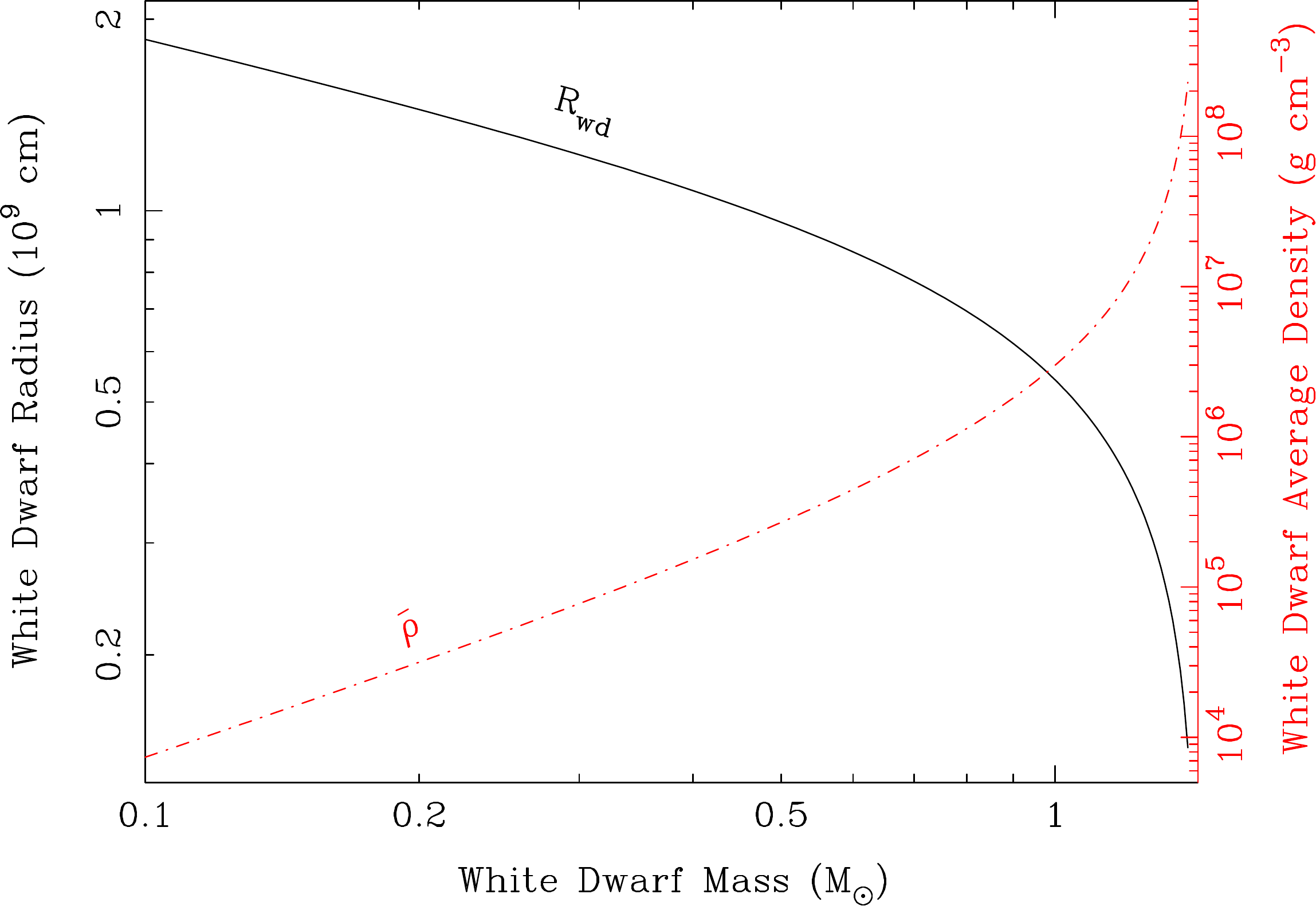}}
    \caption{Radius of a Carbon-Oxygen white dwarf as a function of its mass (in \msun). The mass-radius relation of equation~(\ref{eq:mass_radius}) is adopted here, which is due to \citet{Nauenberg1972}. For reference, the red dot-dashed line (labelled $\bar\rho$ with corresponding scale on the right) shows the average density of the white dwarf as a function of its mass.}
    \label{fig_mass_radius}
\end{figure}

\begin{figure}[t]
    \centerline{\includegraphics[width=0.7\textwidth]{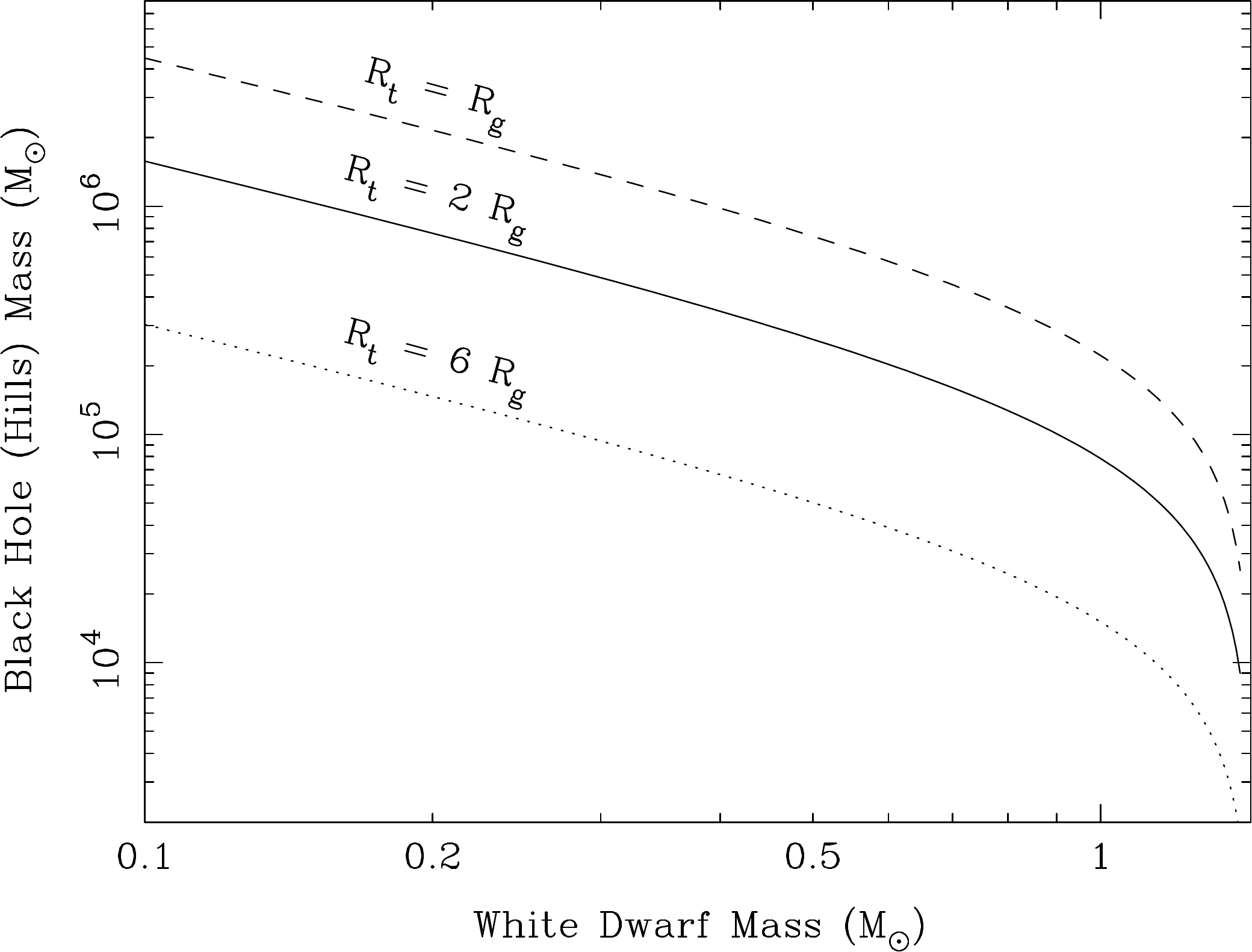}}
    \caption{The variation of the BH Hills mass with white dwarf mass. The three curves represent the boundaries below which the white dwarf is disrupted outside of the innermost stable circular orbit of a non-rotating BH (dotted line, labelled $\Rt=6\Rg$), the event horizon of a non-rotating BH (solid line, labelled $\Rt=2\Rg$) and the event horizon of a maximally-rotating BH, which is also its innermost stable circular orbit (dashed line, labelled $\Rt=\Rg$). The curves are obtained by equating the tidal disruption radius with the appropriate horizon radius and rearranging the equation to express the BH mass in terms of the white dwarf mass.}
    \label{fig_hills}
\end{figure}

White dwarfs are the final products of the evolution of low-mass stars.
They result from stars between $\sim0.1$--8~\Msun that are not massive
enough to ignite further burning stages of their helium or carbon-oxygen
cores. Thus, these cores cool and evolve into very dense stars 
($\sim10^6$~g~cm$^{-3}$) whose masses are determined by their baryon
content, but the pressure necessary to maintain hydrostatic equilibrium 
is provided by the fermionic ``degeneracy pressure" of their electrons. 
While under normal circumstances this is the end of their nuclear
evolution, at least in principle, much more nuclear energy 
can be gained by transforming the white dwarfs (7.08 MeV/nucleon for helium, 7.68 MeV/nucleon for carbon, and 
7.98 MeV/nucleon for oxygen) into iron group elements (8.79 MeV/nucleon):
$\sim 1$ MeV/nucleon or $\sim 2\times 10^{51}$ erg/\Msun can still
be released. Perhaps the most spectacular way to trigger this release, is to send the white dwarf close a black hole (BH) and explosively ignite it by tidal compression \citep{Luminet1989a,rosswog2009ab}.\\
White dwarfs have typical sizes comparable to the Earth's, but following  an inverted mass-radius relation they are smaller the more massive they are, as depicted in Fig.~\ref{fig_mass_radius}. Using a convenient analytic expression derived by \citet[][see also Appendix~A of \citealt{Even2009}]{Nauenberg1972}, we can cast the mass-radius relation of a Carbon-Oxygen white dwarf as 
\begin{equation}
\Rwd=7.80\times 10^8\;
\left(\Mwd\over M_{\rm Ch}\right)^{-1/3}
\left[1-\left(\Mwd\over M_{\rm Ch}\right)^{4/3}\right]^{1/2}
~{\rm cm}
\label{eq:mass_radius}
\end{equation}
where $M_{\rm Ch}=1.435\;\msun$ is the Chandrasekhar mass for that composition. This mass-radius relation, plotted in Fig.~\ref{fig_mass_radius}, is the reason why  the average density of white dwarfs rises sharply  with increasing mass, as shown by the red curve in Fig.~\ref{fig_mass_radius}. \\
BHs are usually classified according to their
masses.
The existence of stellar mass BHs (\lsim 60~\msun) and supermassive BHs ($\sim 10^6$--$10^{10}$ \msun; SMBHs) is well established, but members with masses between these two regimes, so-called intermediate-mass BHs (IMBHs), have not been conclusively identified. They are thought to form via a variety of physical mechanisms, e.g., runaway stellar collisions in a dense star cluster and collapse of a massive gas cloud \citep[see, for example,][]{Portegies2002,Begelman2008,Mayer2010}. \\
To be torn apart by the tidal forces of a BH, a white dwarf must come closer than the tidal radius 
\begin{equation}
\Rt\simeq R_{\rm wd} \left(\frac{\Mbh}{\Mwd}\right)^{1/3}.
\label{eq:Rtid}
\end{equation}
For an accretion disk to form around the BH, the disruption needs to occur outside the innermost stable circular orbit at $R_{\rm ISCO,S}= 6 G \Mbh/c^2\equiv 6R_{\rm g}$ for a Schwarzschild BH or $R_{\rm ISCO,eK}= R_{\rm g}$ for a maximally rotating Kerr BH and a prograde orbit ($\Rg$ is conventionally referred to as the gravitational radius). If the tidal radius lies inside the event horizon ($R_{\rm H,S}= 2 R_{\rm g}$ for the Schwarzschild and $R_{\rm H,eK}= R_{\rm g}$ for the extreme Kerr case), no disruption occurs and the white dwarf is ``swallowed'' whole. These thresholds are plotted as a function of the white white dwarf mass in Fig.~\ref{fig_hills}. Since the tidal radius is proportional to $\Mbh^{1/3}$, but the other radii are proportional to $\Mbh$, there is a (spin-dependent) upper limit on the BH mass beyond which tidal disruption events (TDEs) are impossible. For a non-rotating black hole and a white dwarf of typical mass \cite[$\approx 0.6$ \msun; ][]{Kepler2007}, this limiting BH mass is $\sim 2 \times 10^5$ \msun, therefore only IMBHs can disrupt white dwarfs, see Fig.~\ref{fig_hills}.

\begin{figure}[t]
\centerline{\includegraphics[width=0.75\textwidth]{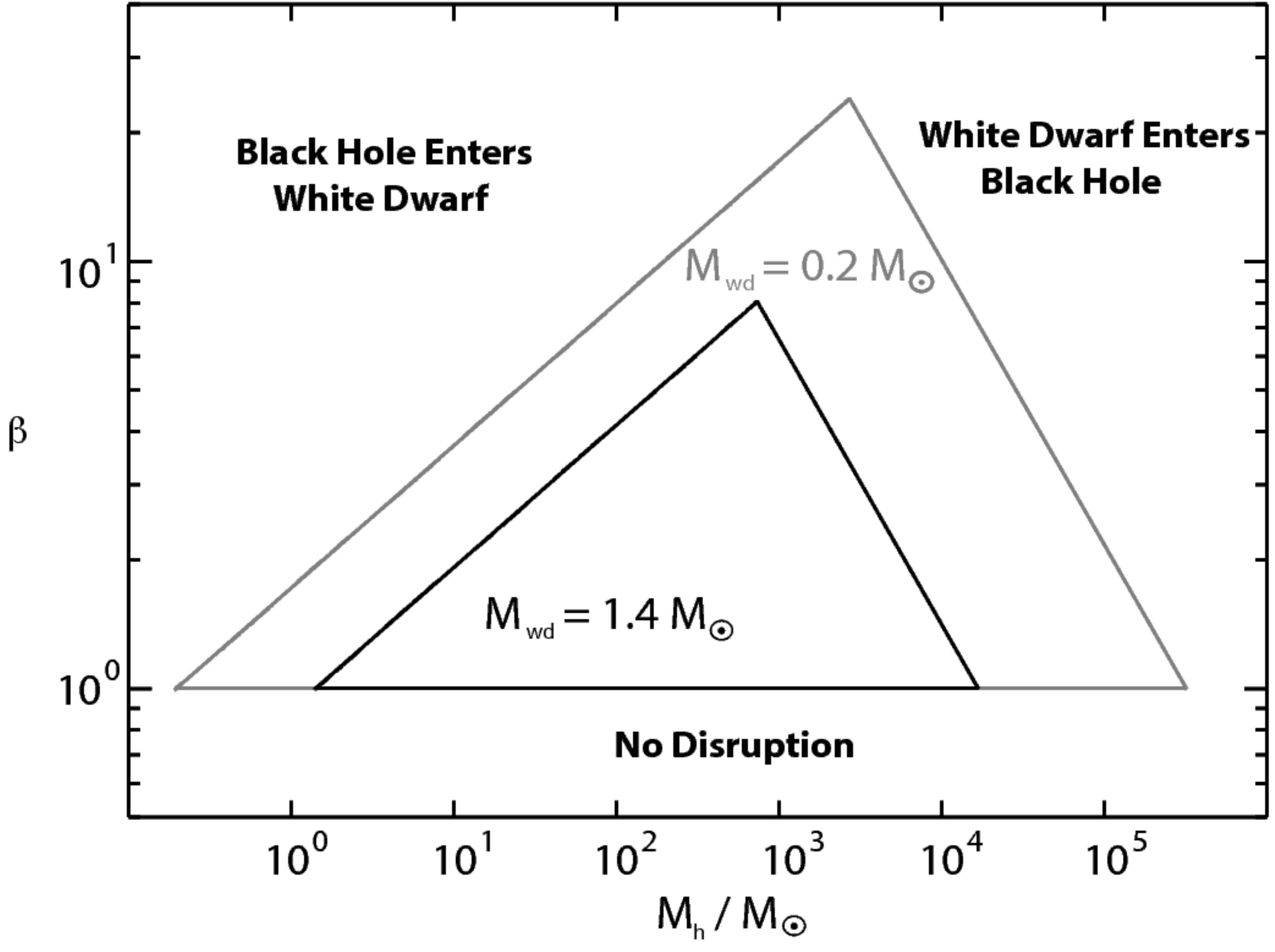}}
\caption{The parameter space defined by the strength of the tidal encounter ($\beta\equiv\Rt/\Rp$) and the BH mass, $M_h$ (assuming a non-rotating BH). This particular figure shows the interesting regions of the diagram for a 1.4~\msun\ and a 0.2~\msun\ white dwarf and was taken from \citet{rosswog2009ab}; an earlier version illustrating the same regions for a 0.6~\msun\ white dwarf was shown by \citet{Luminet1989a}. In the lower region of the diagram, where $\beta < 1$, there is no disruption. If the combination of $\beta$ and $M_h$ falls within the triangle corresponding to the white dwarf's mass, the white dwarf is disrupted outside the event horizon of the BH and the post-disruption debris eventually returns to pericentre. In the upper right corner of the diagram the disruption occurs within the event horizon of the BH. In the upper left corner of the diagram, the black hole event horizon is smaller than the white dwarf, therefore the BH effectively enters the white dwarf during a strong encounter.
}
\label{fig_parmspace}       
\end{figure}

The strength of a tidal encounter is often quantified by the parameter $\beta\equiv\Rt/\Rp$, i.e., the inverse ratio of the pericentre distance to the tidal disruption radius. In other words, a strong encounter is one  where the white dwarf penetrates deeply into the tidal radius. Fig.~\ref{fig_parmspace} illustrates the outcome of a tidal encounter in the $\beta$--$\Mbh$ plane for two different white dwarf masses. For $\beta < 1$ there is no disruption since the white dwarf does not cross the tidal disruption radius. For a given white dwarf mass, the triangle encompasses the region of this parameter space where the debris eventually returns to pericentre after disruption; some of this debris can be subsequently accreted by the BH. For strong encounters outside of this triangle, the BH either enters the white dwarf or it ``swallows'' the white dwarf whole.

The most promising TDE locations depend on the type of disruption.
TDEs of main sequence stars by SMBHs are associated with the cores of galaxies where SMBHs are generally found. However, TDEs involving white dwarfs and IMBHs may occur in a more diverse range of environments since IMBHs are predicted to occur in dwarf galaxies \citep[e.g.,][]{Reines2013}, globular clusters \citep{Jonker2012} and in hyper-compact stellar clusters \citep{Merritt2009b}, although the predicted rates for these different environments differ enormously. 
The discovery of a transient caused by the tidal disruption of a white dwarf by a BH would be clear evidence for the existence of an IMBH. Such a discovery would be extremely interesting for a variety of reasons:
\begin{enumerate}
\item
IMBHs are routinely invoked as ``seeds'' in scenarios for the growth of SMBHs and their co-evolution with their host galaxies \citep[see, for example, the review by][]{Volonteri2012}. Seed masses of at least $10^3\;$\msun\ are required in order to explain the large BH masses found in the highest-redshift quasars \citep[e.g.][ these black holes must have grown to $\sim 10^9\;$\msun\ in $\sim 700\;$Myr by Eddington-limited accretion]{Mortlock2011, Fan2006, Banados2018}.

\item
Dwarf galaxies (with stellar masses $M_* \lsim 10^9\;$\msun) are 
the expected hosts of IMBHs based on an extrapolation of the relation between BH mass and host galaxy stellar spheroid mass \citep[e.g.,][]{McConnell2013} and on the identification of active galactic nuclei (AGNs) in many dwarf galaxies \citep[e.g.,][]{Reines2013,Moran2014,Baldassare2018}. But the occupation fraction, which is critical in setting the white dwarf TDE rate is unknown. Circumstantial arguments by \citet{Silk2017} suggest that an occupation fraction near unity would resolve many of the open questions in dwarf galaxy evolution.

\item
Observations of flares from white dwarf disruptions would afford useful tests of models for the accretion of Hydrogen deficient matter at very high rates relative to the Eddington limit \citep[e.g.,][]{Dai2018}, the formation of jets \citep[e.g.,][]{Krolik2012a}, and the transition from super- to sub-Eddington flows. Since the BH mass is relatively low in this case, the Eddington ratio (accretion rate divided by the Eddington rate) will be much higher than in the disruption of normal stars by more massive BHs and the accretion flow will evolve on a considerably shorter time scale.

\item
The event {\it rates} afford important tests of models for the stellar populations and dynamics in the vicinity of the BH. Since the expected rate estimates depend on a number of assumptions, as we detail later in this chapter, one could potentially eliminate entire families of models based on measurements of the event rates, which would represent substantial progress.

\item
White dwarfs in bound orbits spiraling into IMBHs can produce a detectable gravitational wave signal before disruption \citep[e.g.,][]{Sesana2008}. From this signal we can potentially infer many of the fundamental properties of the system, including the masses of the white dwarf and IMBH, which is extremely useful for testing models of the post-disruption accretion flow (discussed further in Section~\ref{sec_gw}, below, and in other chapters in this volume).

\end{enumerate}

One can appreciate the importance of general relativistic effects during such disruptions by expressing the tidal disruption radius in units of the gravitational radius. Combining equation~(\ref{eq:Rtid}) with the definition of $\beta$ and $\Rg$ one can write
\begin{equation}
{\Rp\over \Rg} = {\Rt\over \beta \Rg} \simeq 
{80\over\beta} \left(R_{\rm WD}\over 10^9\;{\rm cm}\right) 
\left(\Mwd\over 0.6\;\Msun\right)^{-1/3} 
\left(\Mbh\over 10^3\;\Msun\right)^{-2/3} 
\label{eq:RtidRg}.
\end{equation}
Thus, a strong encounter of a typical white dwarf with a $10^4\;\Msun$ BH can have $\Rp/\Rg < 10$, i.e., the white dwarf gets within 5~Schwarzschild radii of the BH during its initial fly-by. Such an encounter is relativistic. 

The aim of this chapter is to discuss the theoretical and observational properties of the transients that result from the tidal disruptions of white dwarfs by IMBHs. In Section 2, the underlying theory and simulations necessary for analysing these interactions are highlighted. In Section 3, the predicted nucleosynthetic yields of systems that result in the detonation of the white dwarf are discussed, while in Section 4 the expected optical and UV signatures of white-dwarf tidal-disruption events (TDEs) are highlighted. In Section 5, observational searches for the tidal disruptions of white dwarfs and candidate events are described. The intrinsic and predicted rates of white dwarf TDEs are detailed in Section 6. A summary of the expected gravitational wave emission from the tidal disruptions of white dwarfs by IMBHs and the cosmic-ray acceleration in ensuing jets is provided in Section~\ref{sec_gw}.

\section{Theory and simulations}
\label{theory}
Some aspects of TDEs can be described to good 
accuracy in a purely analytical way
\citep{Luminet1985a,Luminet1986a,Rees1988a,Stone2013a}, 
but in reality a TDE is a highly 
non-linear interaction between gravity (both self-gravity 
of the star and the ``background" gravity from the BH), 
gas dynamics, radiation, potentially magnetic fields and 
--in extreme cases-- thermonuclear reactions. Therefore, 
most astrophysically relevant questions need to be addressed via careful numerical simulations.

\subsection{Geometrical Challenges}
\begin{figure}[htbp] 
   \centering
    \vspace*{-1cm}
   \centerline{\includegraphics[width=7cm]{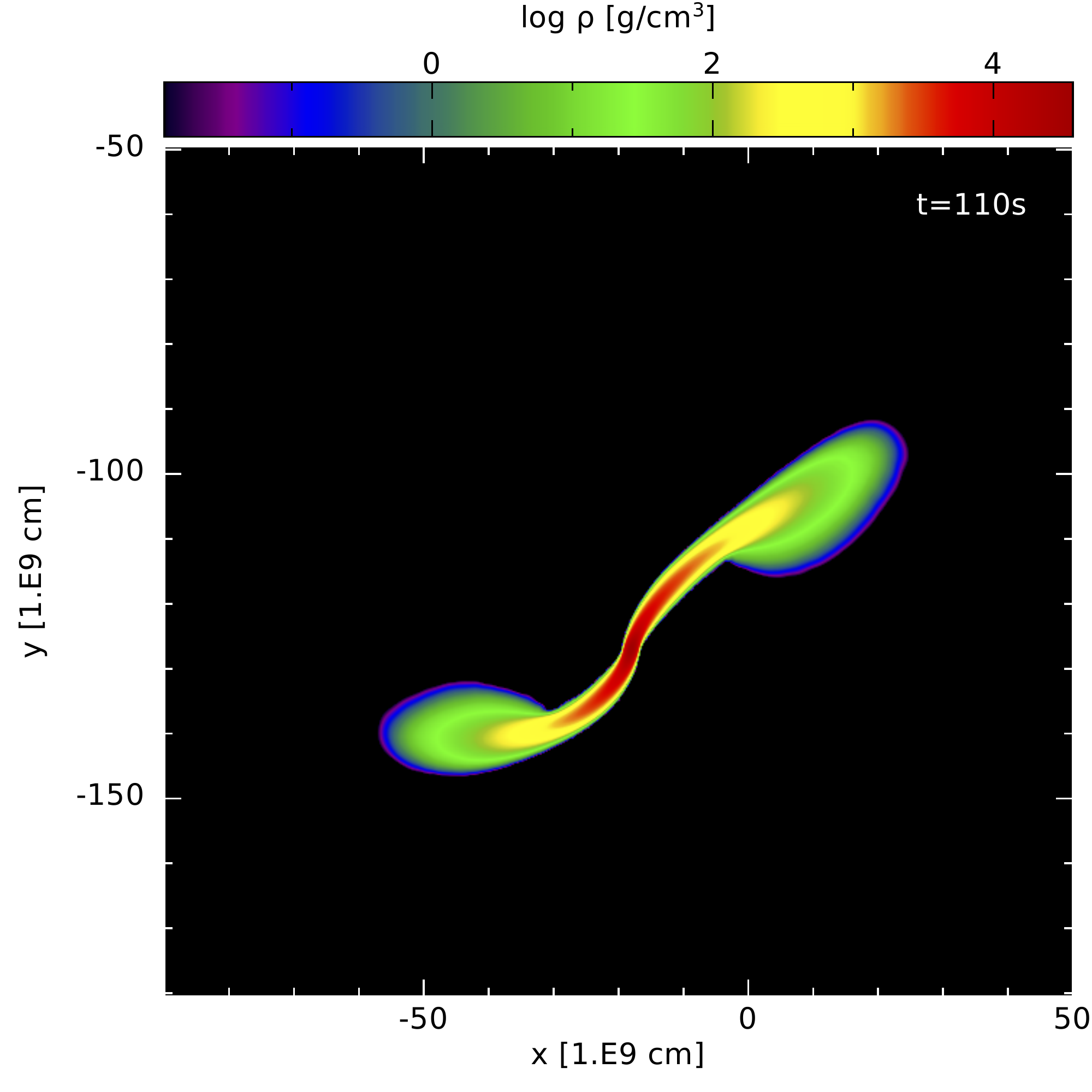} \hspace*{-0.5cm}
               \includegraphics[width=7cm]{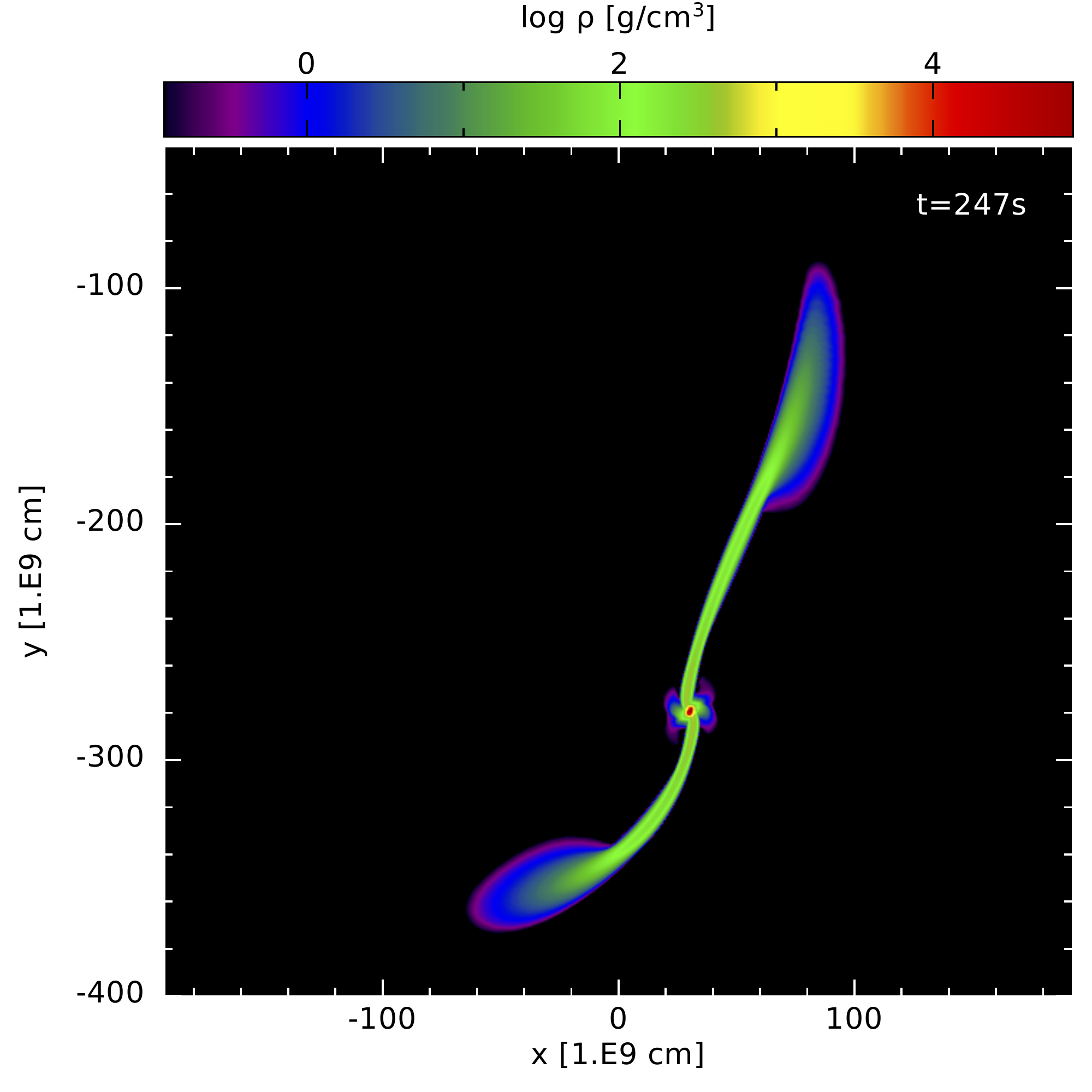}}
   \vspace*{-0cm}
      \caption{Partial tidal disruption of a 0.5 \Msun white dwarf by a 1000 \Msun BH (located at the coordinate origin) illustrating a weak encounter with an impact parameter $\beta= 0.8$ \citep{rosswog19a}.
      The star is ``almost completely'' disrupted (left panel), but a small self-gravitating core survives within the tidal stream connecting the two
      tidal lobes (right panel). The simulation started at an initial separation of $5 R_t$.}
   \label{fig_IMBH_WD}
\end{figure}
The lengths and timescales involved make TDEs formidable
numerical challenges and prevent, in particular 
for most realistic cases, the ``just-do-it" approach, 
where the Einstein and hydrodynamics equations are 
solved together in a consistent way.\\
The initial star far from the BH is in  
hydrostatic equilibrium between gas pressure gradients 
and self-gravity. The acceleration
from self-gravity is $a_{\rm sg}\sim G \Mwd/\Rwd^2$,
while the tidal acceleration at a distance $d$ from the 
BH is $a_{\rm tid}\sim G \Mbh \Rwd/d^3$, so that 
their ratio is 
\begin{equation}
\frac{a_{\rm tid}}{a_{\rm sg}} \sim \left(\frac{\Mbh}{\Mwd}\right)
\left( \frac{\Rwd}{d}\right)^3
\equiv \left(\frac{R_t}{d}\right)^3
\label{eq:acc_ratio},
\end{equation}
which implies, not too surprisingly, a ratio of unity at the tidal radius given by Eq.~(\ref{eq:Rtid}). 
Equation~(\ref{eq:acc_ratio}) shows further that for the tidal
acceleration to be a $<1\%$ perturbation of the
stellar equilibrium, the initial separation, $d_0$, for starting
a numerical simulation needs to be at least
\begin{equation}
d_0 \sim 5 R_t,
\end{equation}
making the initial star a tiny fraction of the simulation
volume:
\begin{equation}
    \left(\frac{\Rwd}{d_0}\right)^3\sim 10^{-8} \left( \frac{\Mwd}{1 \msun}\right) 
    \left( \frac{10^6 \msun}{\Mbh} \right).
    \label{eq:vol_frac}
\end{equation}
This is only dependent on the mass ratio and actually independent 
of the radius of the star that is disrupted. Therefore, the challenge 
is alleviated for an IMBH-white dwarf disruption compared to one where a main
sequence star is disrupted by a SMBH. Eq.~(\ref{eq:vol_frac}) 
has repercussions for simulating TDEs. 
While the small volume covered by the initial star does 
not pose any challenge for a Lagrangian method like smooth particle hydrodynamics 
\citep[SPH;][]{monaghan2005,rosswog2009a,springel2010,rosswog2015a}, 
it is a serious hurdle for Eulerian  methods, where  vacuum has to be treated
as a low-density background gas that 
must be evolved during the simulation.  Therefore, such 
simulations are often performed in the reference frame of 
the stellar centre of mass, with the BH  
being treated as a time-varying, external force \citep{Guillochon2009a,Guillochon2013a}. 
This reduces the computational volume and avoids 
excessive numerical advection error due to high velocity motion 
of the gas with respect to the computational grid.

\subsection{Hydrodynamics and gravity challenges and approaches}
Additional simulation challenges come from 
restriction of the numerical time step. For a full 
space-time solution approach, the relevant signal 
velocity that enters the Courant-Friedrichs-Lewy     
stability criterion \citep{Courant1928} is the speed 
of light, so that the numerical time step is 
restricted to
\begin{equation}
 \Delta t < 10^{-3}  \; {\rm s} \left(\frac{\Delta x} {3 \times 10^7 \; {\rm cm}}\right),
\end{equation}
where $\Delta x$ is the resolution length and its
numerical value for the scaling has been chosen so 
that it corresponds to resolving a typical white 
dwarf radius with $\sim 30$ resolution lengths. 
Self-gravity of the stellar material can play a 
decisive role since it, together with the BH's
tidal field, determines the density structure inside
the star. Since nuclear reactions are very sensitive
to matter density, self-gravity is crucial for an
accurate modeling of a possible explosion.
For weak encounters with $\beta \lsim 1$, also partial disruptions can occur where the star is ``nearly disrupted", but its original stellar core collapses again under its own gravitational pull. Such a situation is illustrated in Fig.~\ref{fig_IMBH_WD} for a 0.5 \Msun white dwarf, a 1000 \Msun BH and a penetration factor $\beta=0.8$. Such surviving cores can have an impact
on the debris properties and subsequent fallback rates
\citep{Guillochon2013a}.
This illustrates the serious challenge of full end-to-end
simulations where a star approaches the BH 
from a large distance, becomes disrupted when passing
the BH into a thin gas stream with a potentially
self-gravitating stellar core, and returns later to 
the BH vicinity to finally settle
into an accretion disk. Thus full space-time solution 
approaches are limited to the near-neighbourhood 
of the BH. \\
To date, the following strategies have been followed
in numerical TDE simulations:
\begin{itemize}
 \item[a)] use an entirely Newtonian approach and restrict the focus to encounters that can be treated as non-relativistic with a reasonable 
 accuracy  \citep[e.g.][]{Guillochon2013a,Guillochon2014a,Coughlin2015,Goicovic2019};
\item[b)] use  Newtonian numerical hydrodynamics  together with a
pseudo-Newtonian potential for approximately capturing some relativistic
effects \citep{rosswog2009ab,Tejeda2013a,Hayasaki2013a,Bonnerot2015a,Gafton2015a};
\item[c)] follow a post-Newtonian approach  for mildly relativistic encounters
\citep{Ayal2000a,Ayal2001a,Hayasaki2016a};
\item[d)] use relativistic hydrodynamics in a fixed BH
space time, but neglect the self-gravity of the star 
\citep[e.g.][]{Anninos2018};
\item[e)] like d), but adding Newtonian self-gravity of the star
\citep{Kobayashi2004a,Rantsiou2008};
\item[f)] use a full numerical relativity approach by 
solving the Einstein equations, and restrict the attention 
mainly to regions near the BH \cite[e.g.][]{Haas2012a,East2014a,Evans2015a});
surviving stellar cores (as seen in Fig.~\ref{fig_IMBH_WD}) in partial disruptions pose 
a very serious challenge to such an approach;
\item[g)] use a combination of some of the above approaches,
e.g. follow the gas motion far from the BH with Newtonian methods and near the BH with relativistic hydrodynamics similar to approach d) \citep[e.g.][]{Shiokawa2015a, Sadowski2016a};
\item[h)] start from a Kerr metric, but include the stellar (white dwarf) contributions to the scalar curvature components in the metric \citep{anninos19} to approximate the stellar self-gravity. This (expensive) correction is switched-off close to the black hole.
\item[i)] use the Newtonian hydrodynamics equations as the computational core, but ``wrap them" into functions that contain the BH metric and its
derivatives \citep{Tejeda2017a,Gafton2019a}; this 
method very accurately reproduces fully relativistic
approaches and includes the self-gravity of the star/its 
debris.
\end{itemize}

\subsection{Black hole tides deforming a nearby star}
\label{sec_deform}
\begin{figure}[htbp] 
   \centering
    \vspace*{-1cm}
   \centerline{\includegraphics[width=12cm]{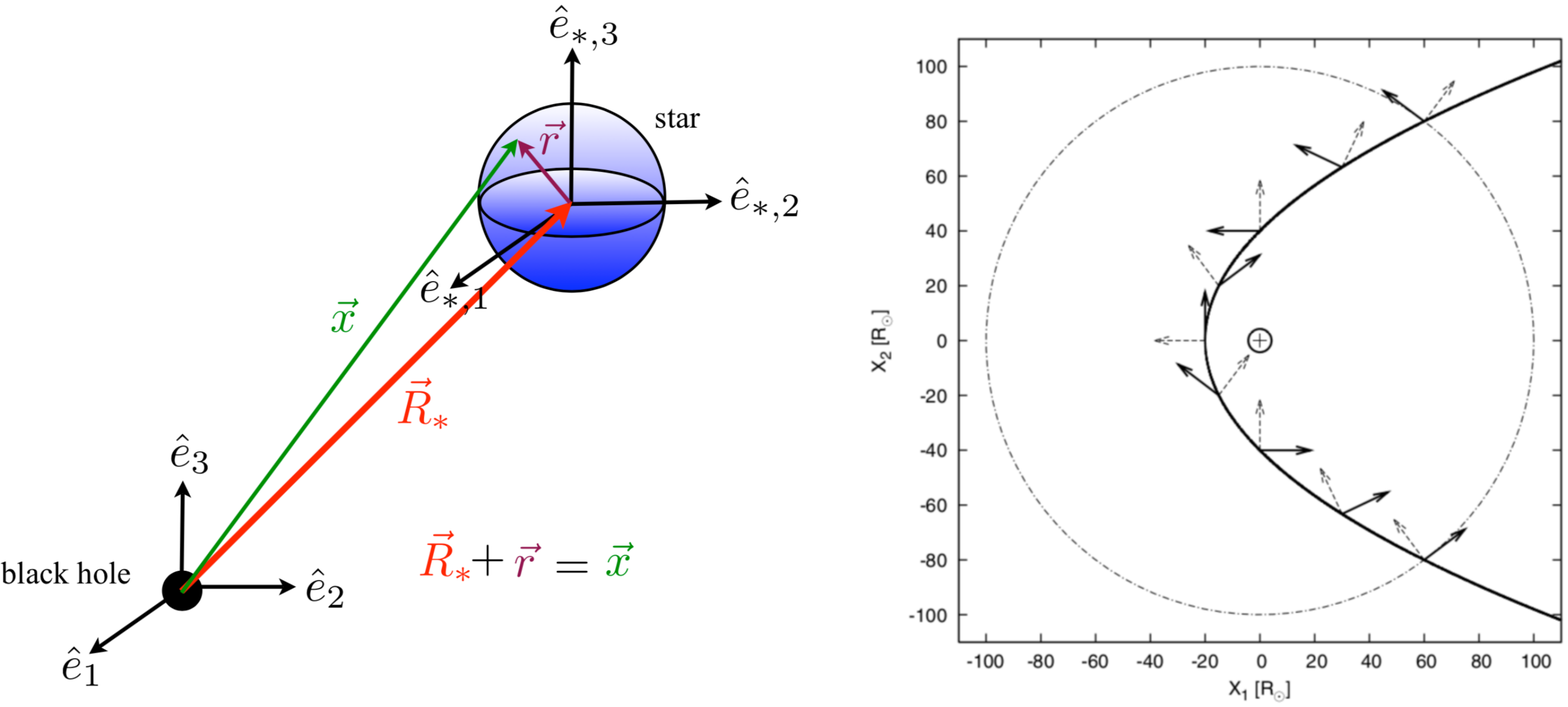}  }
   \vspace*{-1cm}
      \caption{Left: Reference frames
     used for the analysis of the tidal action of a BH on a star with quantities as discussed in Section~\ref{sec_deform}. Right: Eigenvectors
      of the tidal tensor (shown as arrows), taken from \cite{Brassart2008a}. The circle indicates the tidal radius, the parabola the stellar orbit, the BH position is indicated
      by the cross. The third eigendirection (not shown) is perpendicular to the orbital plane.}
   \label{fig_tidal_geometry}
\end{figure}

It is instructive to study how the tidal forces from a BH distort a star.
As a first approximation assume that the BH can be reasonably 
described as a Newtonian point mass. Further assume that a coordinate system 
with unit vectors $\{\hat{e}_1,\hat{e}_2,\hat{e}_3\}$ is attached to the BH and
another coordinate system is centred on the star with unit vectors 
$\{\hat{e}_{\ast,1},\hat{e}_{\ast,2},\hat{e}_{\ast,3}\}$, as sketched in
Fig.~\ref{fig_tidal_geometry}. The center of mass
of the star is displaced from the BH by $\vec{R}_\ast$ and we are interested
in some location inside the star, displaced from the stellar center of mass 
by $\vec{r}$ and from the BH by $\vec{x}= \vec{R}_\ast + \vec{r}$. With these
assumptions, the BH  gravitational potential is
\begin{equation}
\Phi(\vec{x})= - \frac{G \Mbh}{x},
\end{equation}
where $x=|\vec{x}|$. 
For a point  $\vec{x}= \vec{R}_{\ast} + \vec{r}$  inside the star, $\Phi$  can be expanded around $\vec{R}_{\ast}$ into
\begin{eqnarray}
\Phi(\vec{x})&=&  \Phi(\vec{R}_\ast) + 
(\partial_i \Phi)_{\vec{R}_\ast} r_i + 
\frac{1}{2} (\partial_{ij} \Phi)_{\vec{R}_\ast} r_i r_j +
\frac{1}{6} (\partial_{ijk} \Phi)_{\vec{R}_\ast} r_i r_j  r_k + O(r^4)\\
&=& \Phi(\vec{R}_\ast) + 
(\partial_i \Phi)_{\vec{R}_\ast} r_i + \Phi_{\rm tid}(\vec{x}).
\end{eqnarray}
where $\Phi_{\rm tid}$ is the tidal potential.
$\Phi_{\rm tid}$ can be further split up making use of
the tidal tensor (note the sign convention)
\begin{equation}
\tau_{ij}= -(\partial_{ij} \Phi)_{\vec{R}_\ast}
\label{eq:tid_tensor}
\end{equation}
and the deviation tensor
\begin{equation}
\Delta_{ijk}= -(\partial_{ijk} \Phi)_{\vec{R}_\ast}.
\end{equation}
The deviation tensor governs the deviation of the centre of mass
from a pure point mass orbit and it can be shown that
\begin{equation}
M_\ast \frac{d^2 R_{\ast,i}}{dt^2}= - M_\ast \partial_i \Phi(\vec{R}_\ast) + \frac{1}{2} \Delta_{ijk} \int r_j r_k \rho(\vec{r}) d^3r + O(r^3).
\end{equation}
Since $R_\ast \gg r$ the deviations from a point particle orbit are usually negligibly small. The acceleration in the reference frame of the star
is then given by
\begin{equation}
\frac{d^2 r_i}{dt^2}=  \tau_{ij}(\vec{R}_\ast) \; r_j
\label{eq:tidal_accel}
\end{equation}
and it is, therefore, the tidal tensor $\tau_{ij}$ given by Eq.~(\ref{eq:tid_tensor}) that determines how the star
is deformed by the BH. An eigenvector analysis of $\tau_{ij}$ delivers
two negative and one positive eigenvalue, i.e. the star is compressed in two eigendirections
and stretched in the third. Two eigendirections (one positive, one negative) lie in the 
orbital plane and change as the star approaches the BH, see Fig.~\ref{fig_tidal_geometry} (taken from \cite{Brassart2008a}). The third eigendirection,
with negative eigenvalue, is perpendicular to the orbital plane. Thus, as the star approaches the
BH it becomes compressed in the direction perpendicular to the orbital plane.\\
For relativistic BHs, an equation very similar to Eq.~(\ref{eq:tidal_accel}) is found, with
the relativistic tidal tensor, $\tau^{\rm GR}_{ij}$, containing the Riemann curvature tensor. For explicit expressions in
Schwarzschild space-time, see, e.g., \cite{Brassart2010a}, \cite{Cheng2013a}, and \cite{Gafton2015a}.\\

\subsection{Nuclear energy generation}
Triggering dynamically 
important nuclear burning processes during a TDE is non-trivial 
since nuclear reactions are often inefficient and the
compression time during the flyby is very short
 \citep{rosswog2009ab}. An estimate of the compression time, $\tau_{\rm comp}$, can be made assuming the white dwarf is being squeezed with freefall
velocity through a point of maximum compression
(``nozzle") along its orbit:
\begin{equation}
\tau_{\rm comp} \sim \frac{\Rwd}{v_p} \sim 0.2 {\rm s} \left(\frac{\Mwd}{0.6 \msun}\right)^{-1/6}
\left( \frac{\Rwd}{10^9 \rm cm}\right)^{3/2} \left( \frac{\Mbh}{10^3 \msun}\right)^{-1/3}, 
\end{equation}
where the orbital velocity at periastron, $v_{\rm p}$, is
\begin{equation}
v_{\rm p}\sim c \left(\frac{R_{\rm g}}{R_t}\right)^{1/2}\; 
 \simeq 5 \times 10^{9} \; {\rm cm\;s^{-1}} 
 \left(\frac{M_{\rm wd}}{0.6 \msun}\right)^{1/6} 
 \left(\frac{R_{\rm wd}}{10^{9}{\rm cm}}\right)^{-1/2} \left({M_{\rm BH} \over 10^3\;\msun}\right)^{1/3}.
\end{equation}

A simple order of magnitude estimate for the conditions needed to
explode a white dwarf comes from the requirement that the nuclear
energy injection time scale $\tau_{\rm nuc} \equiv \epsilon_{\rm nuc}/\dot{\epsilon}_{\rm nuc}$
must be shorter than the crossing time of the star through periastron, $\tau_{\rm passage}$, so that a substantial amount
of nuclear energy can be released during the short compression time. The burning time scale also has
to be substantially shorter than the dynamical time scale of the star,
$\tau_{\rm dyn}= 1/\sqrt{G \bar{\rho}_{\rm wd}}$, where $\bar{\rho}_{\rm wd}$ is the average white dwarf density, otherwise the star
has time to react on the energy release, can expand and quench the burning. The time to pass the BH at $R_t$, however, is comparable to the dynamical
timescale of the star,
\begin{equation}
   \tau_{\rm passage}\sim \frac{R_t}{v_{\rm p}} \sim \frac{1}{\sqrt{G \Mwd/\Rwd^3}} \sim \tau_{\rm dyn}
   \label{eq:tpass},
\end{equation}
so that both conditions can be summarized as $\tau_{\rm nuc} \ll \tau_{\rm dyn}$.

\subsection{Disk formation}
\begin{figure}[htbp] 
   \centering
   \centerline{\includegraphics[width=6cm]{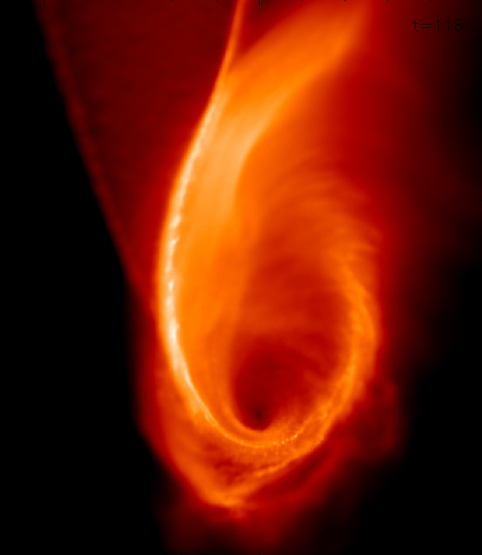}}
      \caption{The debris stream passing the black hole spreads into a "nozzle" of
      large apocentre distance spread. On interacting with the infalling stream a circularization shock can form. Figure from the simulations of \cite{rosswog2009ab}.}
   \label{fig_nozzle}
\end{figure}

The formation of an accretion disk out of the disruption debris is thought to be 
crucial for the electromagnetic appearance of a TDE. This process, however, is to date only incompletely understood, see Chap.~``Formation of an Accretion Flow" of this book. It is, in particular, not obvious that the actual accretion rate, $\dot{M}_{\rm acc}$, should, in all cases, follow the mass delivery/fallback rate, $\dot{M}_{\rm fb} \propto t^{-5/3}$, that is found from simple analytical arguments
and numerical simulations \citep{Rees1988a,phinney89,Evans1989a,Ramirez2009,Lodato2009a,lodato11,Kesden2012a,Cheng2014a}. For this to be true, matter must be accreted faster than it is delivered by fallback. The disk formation process depends on the relative efficiencies of circularization, viscous accretion and radiative cooling \citep{Evans1989a}. All of these topics
currently bear a large degree of uncertainty.
The self-crossing of the accretion stream can be 
efficient in circularizing the disruption debris
into a disk-structure (of some type). It can occur
for at least two reasons. First, the debris that 
passes the black hole has a wide range of specific
energies across the stream width which translates
into a ``fan" spread over a range of apocentre
distances \citep[see][]{Rees1988a,kochanek94}. This fan can collide
and shock with the matter falling towards the 
black hole, as depicted in Fig.~\ref{fig_nozzle}.
Second, if strong, relativistic apsidal precession can lead
to stream self-crossing with enhanced collision angles 
and thus efficiently circularize the debris \citep{Bonnerot2015a,Shiokawa2015a,Hayasaki2016a}. The apsidal precession angle is approximately given by e.g.~\citet{hobson06} and Chap.~``Formation of an Accretion Flow" in this book,
\begin{equation}
\Delta \Phi \approx \frac{3 \pi R_g}{R_p} \approx 11.5^\circ \beta \left(\frac{M_h}{10^6 \msun} \right)^{2/3}  
\label{eq:aps_phi}
\end{equation}
and therefore, is particularly pronounced for high-mass black holes.
\\
\indent \cite{guillochon15} used an approximate scheme with Post-Newtonian orbital prescriptions to 
study the disruption of stars by spinning black 
holes via a Monte Carlo approach. For cases with strong relativistic 
effects and substantial black hole spins, they found that spin-induced 
precession can avoid stream collisions for the times corresponding to
$\sim$10 orbital windings resulting in some cases in substantial 
``dark periods'' directly after the disruptions.\\
\indent They find that for cases where GR effects and precession angles are
small, the streams typically self-intersect far away from the black
holes. This leads to long viscous time scales that can slow down the
accretion rate $\dot{M}_{\rm acc}$ compared to the fallback rate
$\dot{M}_{\rm fb}$
by large factors. Their Monte Carlo studies  suggest
that this is mostly the case for smaller black hole masses. 
Cases with very long viscous time scales
may only peak years after the disruption with bolometric luminosities
decaying more slowly than the $t^{-5/3}$-behaviour that is often used to
observationally identify TDEs. Therefore such events may have gone unnoticed.\\
\indent Specifically since white dwarfs are disrupted by IMBHs, it is currently unclear
if, and in which region of the parameter space, cases with 
very long viscous time scales occur. This is because we have on 
the one hand small apsidal precession angles, see Eq.~(\ref{eq:aps_phi}), 
but on the other hand the low black hole masses favour widely spreading ``fans.'' This interesting question is left for future explorations.

\section{Nucleosynthetic predictions of detonations}
\label{nucleo}

The tidal disruptions of white dwarfs differ from those
of main sequence stars in that there can be an
additional source of energy: the tidal compression 
during the pericentre passage can trigger 
thermonuclear runaway reactions \citep[e.g.][]{Luminet1989a,rosswog2009a,MacLeod2016a,Kawana2017a,Tanikawa2017a,Anninos2018,tanikawa18}. Obviously, it is easier to ignite a He-white dwarf than one with a CO-composition. It should be mentioned, however, that a white dwarf below $\approx 0.46$ \Msun consists of a helium core engulfed in a hydrogen envelope that can extend to several core radii and therefore increase the effective tidal radius \citep{law_smith17}. Alternatively tidal compression could trigger a detonation first in a He-shell on the surface of a white dwarf followed by detonation of a CO core, `tidal double detonation' \citep{2018MNRAS.475L..67T}. 

The first three-dimensional hydrodynamic calculations involving a nuclear network were performed in 2009 \citep{rosswog2009ab} for the tidal disruption of white dwarfs by IMBHs. These simulations used up 
to $4 \times 10^6$ SPH particles and each particle solved a small 
nuclear reaction network \citep{hix98}.
For deep encounters ($\beta>3$) the released nuclear
energy was found to exceed the white dwarf gravitational binding energy and 
to lead to an explosion \citep{rosswog2009ab,Anninos2018}. A 
recent, systematic study (\cite{Gafton2019a}, their Fig. 17; 
though focussing on disruptions of main sequence stars by SMBHs) 
showed that for
$\beta > 3$ a large fraction of the star becomes shocked via the
tidal compression described in Section \ref{sec_deform}, thus 
supporting the idea that the explosion is triggered by compression 
shocks.\\
Reactive flows, however, are notoriously difficult numerical problems, see the excellent review of \citet{mueller98}. The energy release from nuclear reactions creates pressure gradients that trigger fluid motions, which again transport both fuel and ashes to and from the reaction region. In the case of detonations these regions are intrinsically three-dimensional  and one needs to resolve tiny length and time scales.
\cite{Tanikawa2017a} recently suggested that a spatial resolution of $<10^6$ cm (with likely $>10^9$ particles in SPH simulations) is needed to determine if explosive burning will occur. While small nuclear reaction networks can provide an accurate estimate for the nuclear energy release, and thus guarantee a reliable dynamical evolution of the fluid,
substantially larger networks are needed in a post-processing step
to reliably predict nucleosynthetic yields.  

\begin{figure}
\centerline{\includegraphics[width=0.8\textwidth]{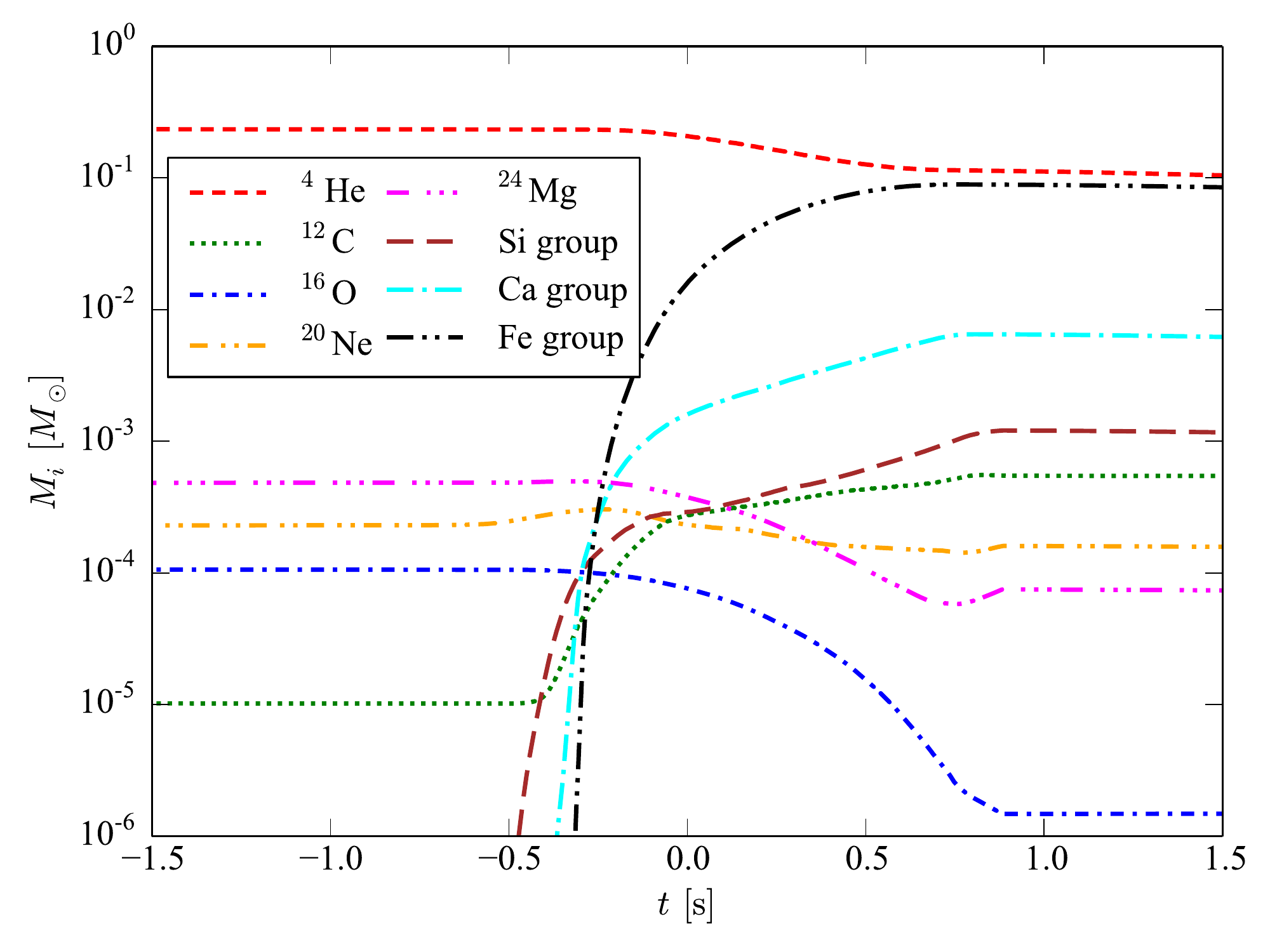}}
\caption{The mass evolution of nuclear species for simulations B3M2R09 (a 0.2 \msun\ white dwarf and  10$^{3}$ \msun\ BH with a $\beta$ parameter of 11). Figure from \citealt{Anninos2018}.  }
\label{fig_anninos_nucleo}       
\end{figure}

The relative yields of intermediate and heavy elements depend
on the density during the burning processes, iron-group 
elements typically require densities $\gsim 10^7$ g cm$^{-3}$.
Such densities can be reached by a combination of the initial
stellar mass (translating into initial density 
and chemical composition), the $\beta$ parameter
(strength of the interaction) and the BH mass \citep{Kawana2017a,Tanikawa2017a,Anninos2018}. In particular
low-mass white dwarfs need very strong encounters to potentially produce iron group elements. Fig.~\ref{fig_anninos_nucleo} shows the
nucleosynthetic production as a function of time for a general 
relativistic calculation of the ignition of a 0.2  \msun\  He white dwarf
by approach to a BH with a mass of 10$^{3}$ \msun\ and 
$\beta$ of 11.  For this model, both Fe-group and intermediate-mass 
elements are produced.
Interestingly, the black hole spin does not have much (if any) impact on the nucleosynthesis \citep{anninos19}.\\
Some observed Ca-rich transients have been suggested to originate from the tidal disruption of white dwarf by a IMBH and it is found that Ca-rich debris can be produced for low $\beta$ systems. These systems, however, also have a low burning efficiency, which results in low yields of intermediate-mass elements and so may not produce enough Ca to be the progenitors of Ca-rich transients \citep{Kawana2017a,Anninos2018}.

\section{Predicted Optical and UV Signatures}\label{sec_ouvsig}

\begin{figure}[t]
\centerline{\includegraphics[width=0.65\textwidth]{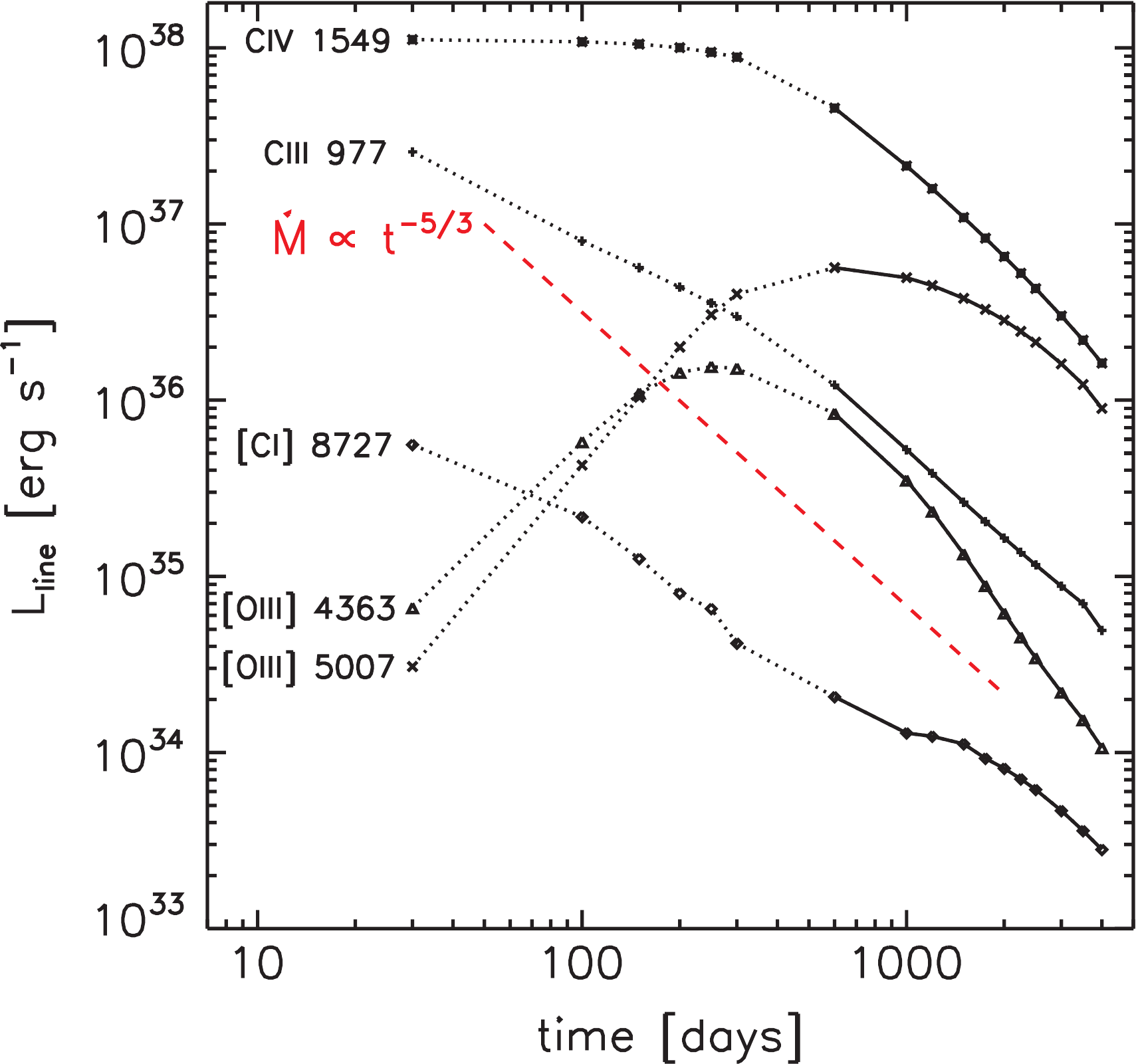}}
\caption{Time evolution of the luminosities of the most prominent emission lines from the photoionized debris produced by the disruption of a 0.55~\msun\ CO white dwarf by a $10^3\;$\msun\ BH in a $\beta=1$ encounter \cite[adapted from][see Section \ref{sec_ouvsig} of the text for the details of the models]{Clausen2011a}. The curves show the evolution of the luminosity of the most prominent emission lines. The dotted portion of each curve (up to 600~days) shows the evolution during and immediately after the super-Eddington accretion phase (440 days for this model), where the model predictions were deemed uncertain. The red, dashed line shows the assumed evolution of the accretion rate which sets the evolution of the strength of the ionizing continuum.
}
\label{fig_lines}       
\end{figure}

The tidal disruption of a white dwarf by an IMBH results in a
number of separate processes that can produce observational
signatures. The disruption of white dwarf material typically 
results in super-Eddington accretion rates which may go
along with jet formation in a manner analogous to the tidal
disruption of main-sequence stars. Unlike main sequence 
stellar disruptions, however, very strong encounters can 
trigger runaway nuclear burning during the first stellar
passage when the star is compressed and shocked as explained
in Section \ref{theory}. This can launch a Type Ia supernova-like event,
but with highly asymmetric outflow.

 The early-time optical and UV spectra are expected to be featureless because of a very large electron scattering opacity in the super-Eddington outflow at this early phase \citep{Strubbe2011a,Roth2016a}. Depending on the geometry and thermal and ionization structure of the outflowing envelope, some emission and/or absorption lines may be discernible in the spectra but these are expected to fall primarily in the far-UV and soft X-ray bands. The models of \citet{Strubbe2011a}, for example, consider the transfer of continuum photons through the outflowing envelope and predict an absorption-line spectrum.  The models of \citet{Roth2016a} consider the random walk of continuum and emission-line photons through the envelope and predict a spectrum with a combination of emission and absorption lines that depends on the particular conditions; some of the easily discernible lines may be in the rest-frame optical part of the spectrum. Although specific calculations of the spectra from the early stages of the disruption of a white dwarf have not been carried out, we can draw an analogy with the work on main-sequence stars described above. Thus we expect the rest-frame optical continuum to be largely featureless while some strong absorption lines or P-Cygni lines from elements that are abundant in a white dwarf will show up in the far-UV portion of the spectrum. 

At later stages of the disruption event, i.e., after the accretion rate drops below the Eddington limit and, presumably, a conventional accretion disk forms, we may expect to see broad emission lines from the tail of returning debris. \citet{Clausen2011a} calculated the optical and near-UV emission-line spectrum from the photoionized debris tail and its evolution over 11~years by adopting and refining the methodology of \citet{Strubbe2009a}. They considered a 0.55~\msun\ CO  white dwarf (67\% O, 32\% C, and 1\% He by mass)\footnote{The mass fraction of H was taken to be $10^{-5}$ and other elements were assumed to have their Solar mass fractions.} disrupted by IMBHs of $10^2$, $10^3$, and $10^4\;\msun$ with encounter strengths of $\beta=1$ and 3.3. They found that, the (permitted) \lion{C}{4}{1550} and \lion{C}{3}{977} \AA\ lines are the two brightest lines in the spectrum and that their strengths decline monotonically with time as the event evolves. During the sub-Eddington accretion rate phase, the strengths of these two lines decline {\it approximately} as $t^{-5/3}$, mirroring the assumed decay rate of the ionizing luminosity. The strongest lines in the rest-frame optical spectra are the (forbidden) \fllion{O}{3}{4959,5007} and \flion{O}{3}{4363} \AA\ lines, although these are always much weaker than the near-UV C lines. This behavior is illustrated in Fig.~\ref{fig_lines}. As this figure shows the O lines initially rise in strength and begin to decline after the super-Eddington phase. The rate of decline of the \flion{O}{3}{5007} line is slower than that of other lines after the first few years. As a result, this line is expected to be the strongest one about a decade after the event. Moreover, since the continuum declines quickly, the observed equivalent width of the line increases in the period from a few years to a decade after the event. \citet{Clausen2011a} also computed the profile of the \flion{O}{3}{5007} \AA\ line and its evolution and they found that, depending on the orientation of the observer relative to the debris tail, the profile is likely to be broad, skewed, and shifted (the shift and width can easily reach thousands of km~s$^{-1}$).

A related family of models by \citet{Clausen2012} examines the emission lines from the debris released in the disruption of the core of an evolved star, specifically a horizontal branch star. After evolving for $\sim10\;$Gyr, the core of a horizontal branch star that started out at 1~\msun, has an enhanced N abundance and suppressed C and O abundances as a result of CNO burning. Because of these modified core abundances and the depletion of hydrogen, the photoionizaiton of the debris leads to strong collisionally-excited emission lines from metals. Specifically, the \flion{O}{3}{5007} and \flion{N}{2}{6583} lines from the debris outshine the Balmer lines about a decade after the tidal disruption. 

\begin{figure}[t]
\begin{center}
\includegraphics[width=\textwidth]{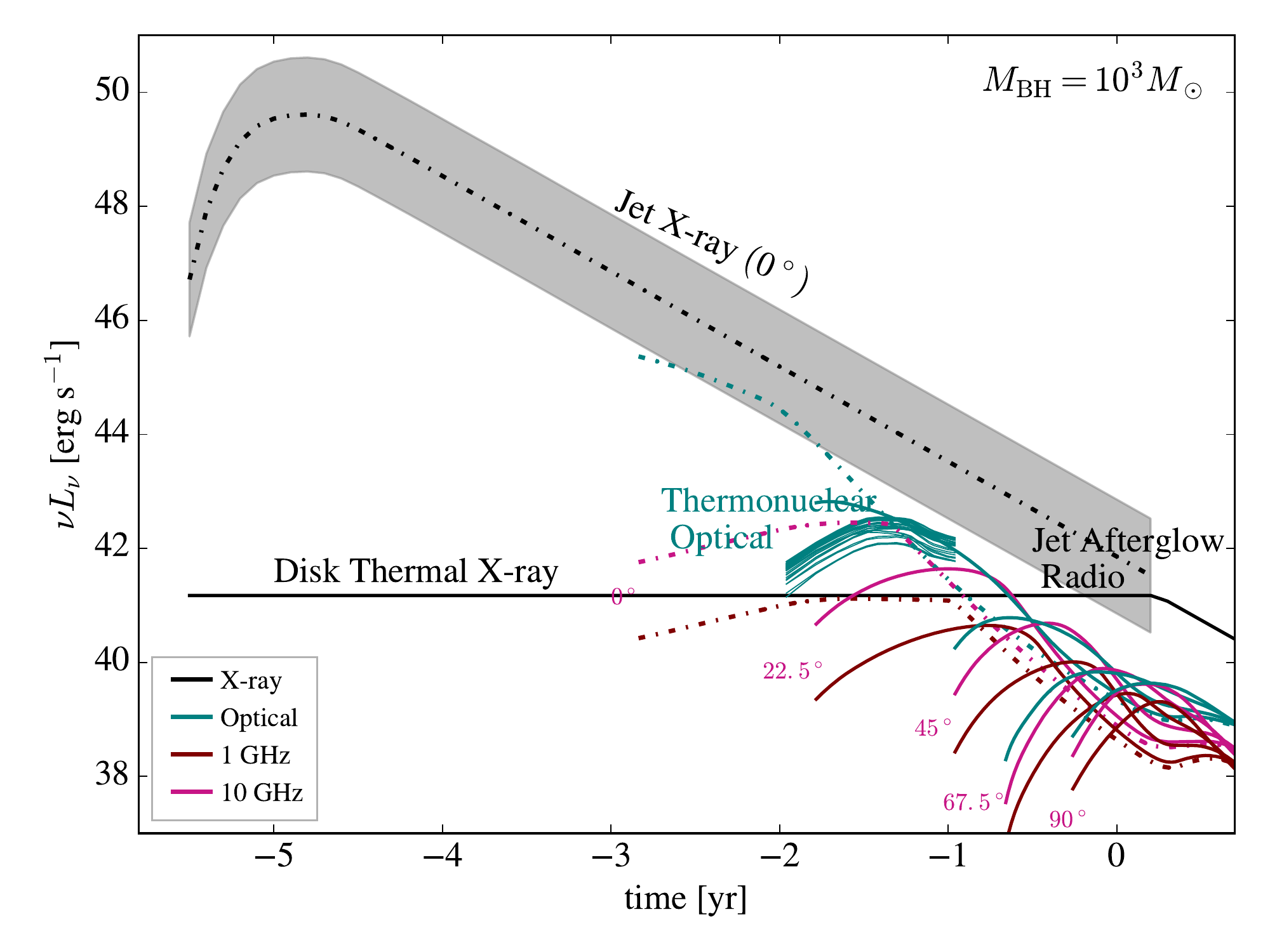}
\caption{The postulated multi-band signatures of a deep-passing white dwarf TDE involving a 10$^3$~\msun\ BH and a 0.6~\msun white dwarf. Times are from time of periapse passage of the white dwarf past the BH.  This figure assumes that a jet is launched carrying kinetic energy of $6\times10^{-3}$ times the accreted rest energy of 0.2~\msun\, or $2\times10^{51}$ erg.  Line styles indicate viewing angle with respect to the jet axis: dot-dashed lines are along the jet axis, while solid lines denote an off-axis observer. Colors show different wavelengths of emission from X-ray to 1 GHz radio. At X-ray and radio wavelengths, the primary signatures are those of the jet and accretion disk. At optical wavelengths, potential signatures include the explosive transient from the thermonuclear burning of the white dwarf, depending on viewing angle, and a possible optical afterglow from the jet. For off-axis events, detection of the thermonuclear transient in the optical is a possible strategy, with accompanying X-ray emission from the accretion disk along with a radio afterglow completing the multi-wavelength picture. Adapted from \cite{MacLeod2016a,2017ApJ...843..154M}.}
\label{macleod_2016}       
\end{center}
\end{figure}

\cite{MacLeod2016a} performed radiative transfer calculations of the tidal compression of a white dwarf during a TDE to obtain predictions of their spectral and light curve properties using the hydrodynamic simulations of \cite{rosswog2009ab} as input. Similarly to Type Ia supernovae, the resulting optical transients are powered by the radioactive decay of $^{56}$Ni produced in the TDE. As discussed in Section \ref{nucleo}, the yield of Fe-group elements will depend on the conditions of the system at the time of disruption, such as the initial white dwarf mass and composition and the $\beta$ parameter. \cite{MacLeod2016a} found that the observed properties (luminosity, optical light curve decline rates, optical spectral features) of the TDEs are highly dependent on the viewing angle but in some cases have properties that are not dissimilar to unusual Type Ia supernovae but are generally less luminous (see examples in Fig.~\ref{macleod_2016}). This thermonuclear transient would dominate over the emission lines, such as the \fllion{O}{3}{4959,5007} and \flion{O}{3}{4363}, produced by the photoionized debris tail discussed above while the transient is at its brightest but at later times these emission lines could become visible \citep{MacLeod2016a}.

\section{Observational searches and candidates}
\label{sec_candidates}

Observational searches for the tidal disruptions of white dwarfs by IMBHs have generally taken two paths, studies looking for fast X-ray transients  and optical searches for Type Ia supernova-like events. In this section we will discuss the properties of observed white dwarf TDE candidates and the link to the theoretical predictions.

\subsection{Fast X-ray candidates}
\label{sec_fastx}

One part of the TDE transient parameter space that is still relatively unexplored is that of fast transients occurring on timescales of minutes or less. At X-ray wavelengths several events with rise times of tens of seconds have recently been discovered. Causality arguments necessitate bright transients on these timescales to be associated with compact sources, and/or jet phenomena. The relatively low BH mass  involved in a tidal disruption of a white dwarf and the compactness of the white dwarf lead to short orbital time scales. Furthermore, as the white dwarf disruption must occur in a strong gravitational field, the circularisation process is probably also short (see the contribution to this Volume by Bonnerot et al.) and as the temperature of an ensuing accretion disc scales as $T\propto M_{BH}^{-0.25}$ (assuming Eddington-limited accretion) the relatively light BHs necessarily involved in white dwarf TDEs could  lead to temperatures high enough for black-body radiation to peak in soft X-rays. Alternatively, given that coronal electron populations responsible for upscattering seed photons to produce hard power-law spectra are thought to be accelerated in magnetic reconnection events (\citealt{merfab2001}), the disruption of magnetic white dwarfs also allows for hard power-law spectra. For all these reasons, white dwarf  TDEs provide appealing explanations to fast X-ray transients.

In the contribution to this Volume from Saxton et al.~the properties of several of the fast X-ray transients discovered are discussed. We here present a brief recap of the properties of these transients investigating if (some of) these are viable white dwarf TDE candidates such as suggested by \cite{2013ApJ...779...14J} and  \cite{2019ApJ...871L..17S}, for example. 

The light curve properties of the observed fast X-ray events are not uniform. The rise and decay times, observed fluxes and fluences show significant diversity. For example, the main peak(s) of the flares varied in duration between $\sim$50--700 seconds. Furthermore, for one event (XRT~000519; \citealt{2013ApJ...779...14J}) two precursor flares were found, each 4000~s before the previous flare. The main peak of the flare was also split in two parts with significant spectral softening between the two peaks. Such precursor events were not detected in any of the other candidates but this may be explained by differences in orbital parameters of the disrupted white dwarfs (e.g.~stars on eccentric orbits providing TDEs with precursor events and single-passage stars on parabolic orbits providing one-off flares; \citealt{MacLeod2014b}). The timescale of the precursor flares is in line with the expected orbital timescale of a white dwarf in (an eccentric) orbit around an IMBH (cf.~Fig.~5 in \citealt{2016ApJ...819...70M}). Unfortunately, the distances to the sources responsible for these fast X-ray flares are in most cases not well constrained as no redshift information is available. However, for two flare events, there was spatial coincidence with Virgo cluster galaxies, making it probable that these flares occurred in the outskirts of these systems \citep{2013ApJ...779...14J,2016Natur.538..356I}. The luminosities calculated assuming galaxy associations are consistent with the Eddington limit for IMBHs.

The X-ray spectral shape of these fast events is consistent with a power law (see the discussion and references in the contribution of Saxton et al.~to this Volume). This is in contrast to the early-time soft X-ray spectra typically found in TDEs involving more massive black holes and stars other than white dwarfs (for some references and a discussion on differences between early- and late-time X-ray spectra of TDEs see for instance \citealt{jonker2019}). It seems unlikely that a difference in the (Eddington ratio) mass accretion rate in white dwarf and main-sequence TDEs is the main cause of the different spectral shape\footnote{Although the process of circularisation of the gas stream in white dwarf disruptions is affected by General Relativistic effects in contrast to many main sequence TDEs; see the contribution from Bonnerot et al.~in this Volume on the formation of the accretion flow after disruption}. A probable difference between main-sequence and white dwarf TDEs is the magnetic field of the star before disruption. The typically weak magnetic fields of main sequence stars may imply that their TDE discs start off with low levels of magnetic fields, whereas several white dwarf types host significant magnetic fields implying that accretion flows from white dwarf TDEs may start off with a higher magnetic field. Since power-law X-ray spectra are thought to be caused by Compton up-scattering of seed disc photons by a coronal electron population, which in turn is thought to be accelerated to relativistic energies in magnetic reconnection events (\citealt{merfab2001}), a power-law spectrum may more naturally emerge in the TDE discs with large magnetisation such as those occurring in white dwarf disruptions. Note that a main-sequence/longer duration TDEs with hard power law like spectra are also more difficult to distinguish from AGN activity and therefore observers may be biased against classifying such events as TDEs (\citealt{jonker2019}).

The extreme source Swift~J164449.3+573451 has been studied in detail, warranting a separate paragraph: it was discovered at the centre of a compact galaxy at $z=0.35$ \citep{Levan2011} due to its $\gamma$-ray emission. It displayed $\gamma$-ray emission lasting a few days (much longer than the duration of seconds to minutes for typical gamma-ray bursts, GRBs), as well as an extremely long-lived X-ray component. A number of models have been suggested to explain this event including a relativistic jet launched due to a TDE \citep{Bloom2011} and the formation of an accretion disk after the collapse of a massive star to a BH (\citealt{2012MNRAS.419L...1Q}; \citealt{2012ApJ...752...32W}). \cite{2011ApJ...743..134K} suggested that jets from the tidal disruption of a white dwarf by an IMBH can explain the nature of this transient, in particular the short-term variability in and the fast rise time of its X-ray light curve. However, even though \citet{2014MNRAS.437.2744T} do not rule out that this event was powered by the tidal disruption of a white dwarf, this scenario is disfavoured according to their modelling.

There is some overlap in the X-ray properties of classical GRBs and the X-ray transients described above. Some X-ray transients have been interpreted as being caused by a GRB observed off-axis (e.g.~\citealt{Urata_2015}), as well as stripped envelope supernovae (\citealt{2005ApJ...627..877S}). Conversely, several ultra-long GRBs have been suggested to be associated with white dwarf TDEs (e.g.~\citealt{2014ApJ...781...13L}), although other explanations have been proposed as well (e.g.~\citealt{2016ApJ...823..113P}; \citealt{2010ApJ...717..268G}). In order to check the potential association, at least statistically, we compare the observed and predicted event rates.

The rate of Type Ibc supernovae (some of which are associated with GRBs) is 0.258$\times$10$^{-4}$ Mpc$^{-3}$ yr$^{-1}$ and if we include the estimate from \citet{2004ApJ...607L..13S} that at most 6 per cent of Type Ibc supernovae produce a collimated jet this yields a rate of 1.55$\times$10$^{-6}$ Type Ibc supernovae with collimated (GRB) jets per Mpc$^{-3}$ yr$^{-1}$. For an assumed typical X-ray luminosity of L$_X\sim$10$^{48}$erg s$^{-1}$ for the X-ray flash associated with the GRB, a distance limit of nearly $\sim$10 Gpc is implied given the observed X-ray fluxes of $\sim10^{-10}$ erg cm$^{-2}$ s$^{-1}$ of the fast X-ray transients. Using this we derive that the Type Ibc supernova/GRB rate in this volume will be $\sim$1.5$\times$10$^{6}$ yr$^{-1}$. So, comparing the observed X-ray flash rate of 1.4$\times 10^5$ yr$^{-1}$ over the whole sky as calculated by \citet{2015MNRAS.450.3765G} with these predicted rates, we conclude that the X-ray flashes discussed in the contribution of Saxton et al. to this Volume could be due to GRBs associated with Type Ibc supernovae. We do note that the uncertainties in these rate calculations are still substantial for instance if only due to Poisson fluctuations given the low number of detected fast X-ray flashes.

In Section~\ref{sec_rates} we estimate the rate of X-ray flares expected from white dwarf TDEs. The end result of those deliberations on the rate $R$ within the volume probed with a flux limit set to $\sim10^{-10}$ erg cm$^{-2}$ s$^{-1}$ is $R=f_{occ} \times 10^4$ yr$^{-1}$. This value falls a factor of about $14/f_{occ}$ short of the observed rate of X-ray flashes of 1.4$\times 10^5$ yr$^{-1}$. Of course, as indicated in Section~\ref{sec_rates}, there are significant uncertainties in these rate calculations. Nevertheless, even optimistically assuming an occupation fraction of 1 leaves a large gap between the predictions of X-ray flares from white dwarf TDEs and the observed rate of X-ray flares. Contributions from additional populations of IMBHs, such as those that have been speculated to reside in globular clusters, or hyper--compact stellar clusters, or the lighter BHs in the recently--discovered stripped nuclei (\citealt{fragione2018}; \citealt{merritt2009}; \citealt{voggel2019}, respectively), may increase the rate somewhat. However, it is likely that only a small fraction of globular clusters host IMBHs and the feeding rate of white dwarfs into the loss cone is likely too low for the majority of these objects (see again also Section~\ref{sec_rates} for details). A high spin rate that increases the Hills mass for BHs with masses $<10^6$\Msun\  could enable IMBHs to potentially disrupt more white dwarfs but overall, it is difficult to explain all the X-ray flares as solely due to white dwarf TDEs.

\subsection{Optical candidates} 

As discussed in Section \ref{sec_ouvsig}, if the white
dwarf is tidally compressed and ignites nuclear burning 
then the dominant powering mechanism of optical 
transients in the days to weeks after disruption will be the radioactive decay of Fe-group elements,
provided they are produced in sufficient amounts
(e.g. in very strong encounters and/or massive white dwarfs). Each Type Ia supernova typically produces 0.6 \msun\ of Fe-group elements per explosion. MacLeod et al. (2016) investigated the nucleosynthetic yields of white dwarf+IMBH TDEs and found that the amount of Fe-group elements synthesised may be significantly lower than that seen in normal Type Ia supernovae, which affects their observed light curves and spectral properties.

\textit{Ca-rich transients} are a class of observed optical transients with magnitudes of $\sim$ $-$16 (fainter than normal Type Ia supernovae) that have strong features of Ca in their late-time spectra, suggesting that they may have appreciable Ca in their ejecta  \citep{2010Natur.465..322P, 2012ApJ...755..161K,2017ApJ...836...60L}. Caution in associating strong Ca features with large quantities of Ca, however, should be exercised because Ca is a very efficient coolant and a small amount of Ca in the ejecta can produce a strong feature. The early-time spectra of Ca-rich events have optical spectra that look similar to stripped envelope Type Ib supernovae that typically show strong He lines  \citep[e.g.][]{2010Natur.465..322P}. Various models to explain  Ca-rich transients have been suggested including TDEs involving a white dwarf disrupted by an IMBH \citep{2015MNRAS.450.4198S}, a He-shell detonation on the surface of a white dwarf \citep[`.Ia supernova' ][]{2010Natur.465..322P,2010ApJ...715..767S}, and collisions involving a He donor and a CO or ONe white dwarf \citep{2017MNRAS.468.4815G}. The original white dwarf TDE models are from \cite{rosswog2009a} but radiative transfer calculations were performed by \cite{MacLeod2016a} and some low-mass white dwarf disruptions may result in transients rich in Ca. The key distinguishing characteristic for determining if Ca-rich transients are linked to TDEs involving white dwarfs and IMBHs is the detection of high-energy emission. Supernovae originating from the thermonuclear explosion of white dwarfs have not been detected in the X-ray band but X-ray emission is a key prediction of TDE models. A search for X-ray emission was made for SN 2016hnk, a white dwarf TDE candidate, but only an upper limit was placed \citep{2018MNRAS.475L.111S} but future searches in the X-ray band will be important for testing the potential link between Ca-rich transients and white dwarf TDEs.

\subsection{Multi-wavelength candidates}

Some candidate white dwarf and IMBH TDEs have been discovered that are bright in the optical as well as having high-energy counterparts.

\begin{figure}
\centerline{\includegraphics[angle=-90,width=0.75\textwidth]{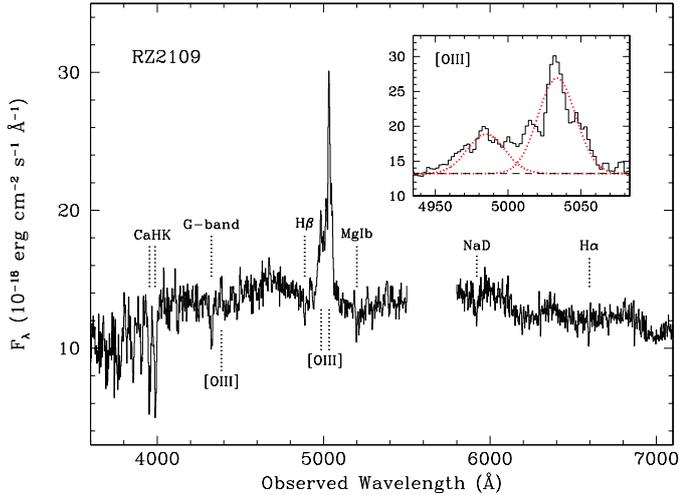}}
\caption{The optical spectrum of the globular cluster RZ~2109 associated with the elliptical galaxy NGC~4472 in Virgo, adapted from \cite{Zepf2008}. The dotted vertical lines and labels {\it above} the continuum identify some of the prominent absorption lines in the spectrum of an old stellar population. The dotted vertical lines and labels {\it below} the continuum identify three of the stronger \fion{O}{3} lines expected in this range. The inset shows the profile of the \fion{O}{3}{4959,5007} lines with Gaussian models of FWHM$\,\sim\,$2000~km~s$^{-1}$ superposed.}
\label{fig_rz}       
\end{figure}

\begin{description}

\item[\textit{AT2018cow:}] A rapidly rising and luminous transient was discovered in the optical by \cite{2018ApJ...865L...3P} and also detected in the X-ray, radio, and submillimeter  \citep{2018MNRAS.480L.146R, 2019MNRAS.tmp...53K,2018arXiv181010720M}. A TDE involving a He-star and an IMBH was suggested by \cite{2019ApJ...871...73H}. \cite{2019MNRAS.484.1031P} suggested that the transient could be caused by a main-sequence star disrupted by an IMBH but this is at odds with the detection of dense circumstellar material (detected via radio measurements) that would need to be ejected from the system prior to the TDE \citep{2018arXiv181010720M}. Other suggestions for the origin of this transient include the core-collapse of a low-mass He star or a stellar collapse involving accretion onto a newly formed compact object \citep{2018ApJ...865L...3P,2018arXiv181010720M}.

\item[\textit{RZ~2109:}] \citet{Maccarone2007} reported the discovery of an accreting BH in the globular cluster RZ~2109, associated with the elliptical galaxy NGC~4472 in the Virgo cluster, based on its X-ray luminosity and variability properties. Followup optical spectroscopy of this object by \citet{Zepf2008} revealed the remarkable spectrum illustrated in Fig.~\ref{fig_rz}, which comprises a continuum from the old stellar population of the globular cluster and a very broad emission feature identified with the \fllion{O}{3}{4959,5007} doublet \citep[see also][]{Steele2011}. The lack of any other optical emission lines prompted \citet{Clausen2011a} to propose that the properties of this system were the result of the tidal disruption of a CO white dwarf by an IMBH at the centre of the cluster, accretion of some of the debris, and the photoionization of remaining, returning debris by accretion-powered ionizing radiation. This proposal was bolstered by photoionization calculations that explained the emission-line spectrum (see discussion of models in Section~\ref{sec_ouvsig}) and a model for the profile of the \fllion{O}{3}{5007} line that reproduced the observed profile. 

However, alternative explanations were also proposed, including accretion of hydrogen-deficient material from a red giant by the BH \citep{Porter2010}, emission from the wind of an R Corona Borealis star \citep{Maccarone2011}, and the ionization of the ejecta of a nearby nova by the accreting BH \citep{Ripamonti2012}. Moreover, some of the observed properties of RZ~2109 cannot be explained by the TDE scenario, e.g., the fact that the X-ray luminosity has remained relatively steady for about 16 years \citep{Dage2018} and the fact that the source of the emission lines is likely to be extended \citep{Peacock2012}. Thus, the nature of the X-ray and emission-line sources in RZ~2109 remains ambiguous. Another globular cluster with properties reminiscent of RZ~2109 was found in the elliptical galaxy NGC~1399 by \citet[][it is associated with the X-ray source CXO~J033831.8$-$352604]{Irwin2010}. The optical spectrum of this object features the \fllion{O}{3}{4959,5007} doublet as well as the \fllion{N}{2}{6548,6584} doublet but no other strong emission lines, which inspired \citet{Clausen2012} to explain it by invoking the disruption of a blue horizontal branch star. But all the alternative models proposed for RZ~2109 apply to this object as well. 

\end{description}

\section{Rates: intrinsic and predicted in surveys}
\label{sec_rates}
White dwarf TDEs, and their associated flares, are likely rare compared to TDEs of main sequence stars. There are several reasons for this. First, the compactness of white dwarfs implies that a more extreme tidal force is needed to disrupt them than is needed for a main sequence star. This implies a closer periapse approach to the BH is needed, and there is a correspondingly smaller cross section for these interactions. Further, only IMBHs with mass $\Mbh \lsim 10^5 \msun$, can accomodate these close periapse approaches without swallowing the white dwarf whole. Second, white dwarfs, while not rare in an old stellar population, are less common than low-mass main sequence stars. Nonetheless, the potentially luminous and distinct signatures of white dwarf disruption may separate these events from the larger number of main-sequence disruptions.\\

\subsection{White dwarf tidal disruption sites and dynamics}
\label{sec_dynamics}

Just like stars of other stellar-evolutionary states, white dwarfs are tidally disrupted after passing at very close periapse distance to a massive BH. These events rely on the coexistence of massive BHs and surrounding dense clusters of stars. While highly uncertain at the BH masses most relevant to white dwarf tidal disruptions, in more massive BHs, these clusters are thought to contain roughly a BH mass of stars in a roughly isotropic, dynamically relaxed cluster surrounding the BH. Within this cluster, stars trace wandering orbits directed by both the gravitational attraction of the BH and interaction with all of the other stars. These wandering orbits occasionally are scattered to very high eccentricity, which causes them to plunge to close approaches to the BH (see the contribution of Stone et al. to this Volume on the processes that  send stars on orbits that will lead to their tidal disruption). 

Currently, predictions for the occurrence of white dwarf tidal disruptions come from two distinct methods: scaled-down models of stellar cusps present around more massive BHs, and N-body models of globular clusters containing IMBHs of hundreds to thousands of solar masses. We begin by examining the qualitative predictions of the nuclear cluster models. These models are discussed in detail by \citet[][chapter 6]{2013degn.book.....M} and in the contribution of Stone et al. to this volume; they are applied to the white dwarf tidal disruption context by \citet{MacLeod2014b,MacLeod2016a}. In such clusters, a region of similar size to the BH sphere of gravitational influence $r_{\rm h} = G \Mbh/ \sigma^2$, where $\sigma$ is the velocity dispersion of the surrounding stellar distribution, is assumed to contain approximately a BH mass of stars, arranged with a power-law profile in number density as a function of radius \citep{Bahcall1976,Bahcall1977, Frank1976}. 
In the discussion that follows, we make the  simplifying assumption that all stellar subtypes are distributed homogeneously within this power law profile. However, there are several reasons that this simplification may not occur in reality. Over a cluster relaxation time, stars of different mass segregate such that more massive objects end up more tightly bound to the BH than less tightly bound objects \citep[See chapter 7 of][]{2013degn.book.....M}. This effect can be particularly strong if there is a significant population of stellar mass BHs, with mass much greater than the average cluster star \citep{2009ApJ...697.1861A,2016ApJ...830L...1A}.

In scaling the properties of these clusters down from those of more massive BHs, \citet{MacLeod2014b} applied the BH mass-velocity dispersion relation of \citet{Kormendy2013}, we note here that this has the consequence of determining the, a priori uncertain, density of the cluster that surrounds the putative BH. This point of uncertainty merits some caution, because, as we describe below, typical stellar densities in globular clusters (which could potentially harbor black holes of less than approximately $10^3M_\odot$) have sufficiently low binding energy that the presence of a black hole modifies the entire cluster distribution, not just a central cusp. 

\begin{figure}[tbp]
\begin{center}
\includegraphics[width=0.85\textwidth]{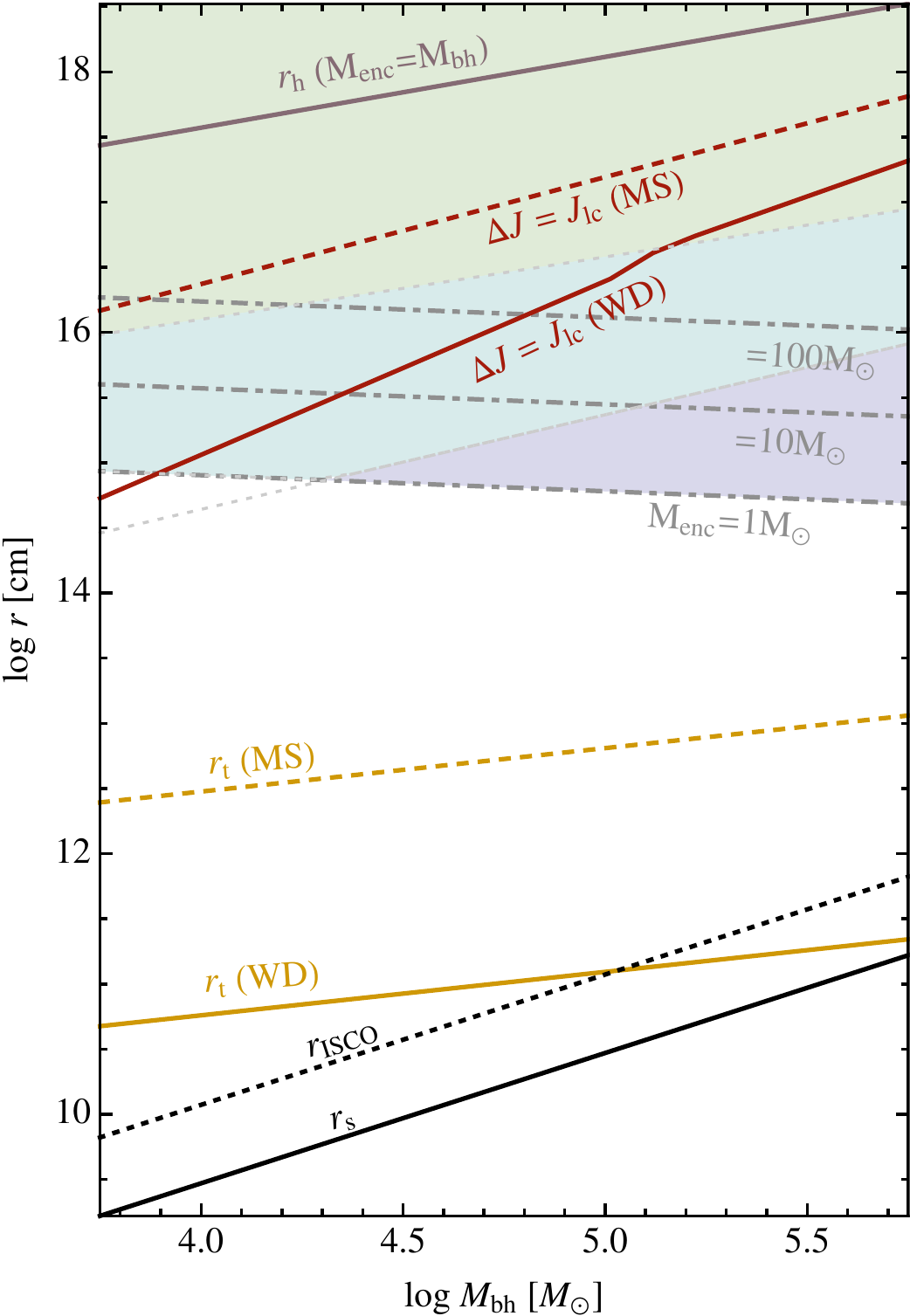}
\caption{Characteristic scales for interactions of white dwarfs within a central stellar cluster with the central massive BH, as a function of varying BH mass. Here the cluster is assumed to contain one BH mass of stars within the BH's sphere of gravitational influence, arranged as $n(r) \propto r^{-1.5}$. Shown, from bottom to top, are: 1) the Schwarzschild radius, $r_s$, and the radius of the Innermost Stable Circular Orbit $\sim 4 r_s$ (black solid and dotted, respectively), 2) the tidal radius, $r_t$ for white dwarfs $(0.5M_\odot)$ and MS (sun-like) stars (yellow solid and dashed), 3) the radii that enclose 1, 10, and 100 $M_\odot$ (gray dot-dashed), 4) the characteristic orbital semi-major axis that marks the transition from the empty (smaller $a$) to the full (larger $a$) loss cone regimes (red solid and dashed lines labeled $\Delta J = J_{lc}$), and 5) the MBH radius of influence, $r_{\rm h}$. Filled regions are, on average, populated with stars. Filling colors denote the primary orbital relaxation mechanism with general-relativistic resonant relaxation (purple), mass-precession resonant relaxation (cyan), and finally non-resonant relaxation (green) being dominant from small to large radii, respectively. Figure adapted from \citet{MacLeod2014b}.}
\label{macleod_2014}       
\end{center}
\end{figure}

With stellar cluster properties defined, one can compute the various ``relaxation" properties that cause stellar orbits to change over time. In general, these include the stochastic scatterings of two-body gravitational encounters (two-body relaxation), and the coherent, secular torques applied to closer-in stellar orbits over many orbital cycles \citep[resonant relaxation; see][]{2013degn.book.....M}. The typical magnitude of the per orbit random-walk in angular momentum space can be expressed in terms of the relevant relaxation timescale, $t_{\rm rel}$ as $\Delta J \approx J_{\rm c} \left( P/ t_{\rm rel}\right)^{1/2}$, where $P$ is the orbital period and $J_{\rm c}\approx \sqrt{G M_{\rm bh} a}$ is the angular momentum of a circular orbit. This expression implies that over one relaxation time the orbital angular momentum changes by of order the circular angular momentum.  Fig.~\ref{macleod_2014} shows the characteristic radii in such a nuclear star cluster as a function of BH mass, adapted from \citet{MacLeod2014b}. The line $\Delta J = J_{\rm lc}(WD)$ marks the orbital semi-major axis within the cluster where the root mean square change in orbital angular momentum over one orbit, $\Delta J$, equals the angular momentum of a ``loss cone" orbit, $J_{\rm lc}(WD)$, that has a  periapse distance similar to the white dwarf tidal radius. 

At radii larger than this line, the low angular momentum phase space depleted by white dwarf tidal disruptions is repopulated each orbital period (and as a result, tidal encounters of all impact parameters occur). This is the ``full loss cone" portion of the cluster phase space. At radii smaller than the $\Delta J = J_{\rm lc}(WD)$ line, the per orbit scatter is smaller than the loss cone angular momentum. This implies that the loss cone phase space is ``empty" and that only grazing impact parameter encounters occur. \citet[][section 3]{MacLeod2014b} showed that most white dwarf TDEs are fed from regions where the loss cone is efficiently refilled by either two-body relaxation or mass-precession resonant relaxation. Under the assumptions of this representative cluster model, white dwarf tidal disruptions therefore occur primarily in the full loss cone limit, and therefore distribute across all periapse impact parameters. 

The overall rate of white dwarf TDEs in such a model was found to be approximately $10^{-6}$~yr$^{-1}$ per BH. For comparison main-sequence star TDEs occurred with a frequency of $10^{-4}$~yr$^{-1}$ per BH in these same models. The white dwarf tidal disruption rate was therefore 1\% of the overall tidal disruption occurrence rate. Although many assumptions must be made about the properties of stellar clusters surrounding BHs to estimate this rate,  \citet{MacLeod2014b,MacLeod2016a} have argued that the ratio of white dwarf to main sequence tidal disruptions may  be robust with respect to variations in the system properties. Within a model in which stars are distributed in a homogenous, power-law profile, the unknown parameter with the strongest effect on the relative rates of main sequence versus white dwarf disruption events is the slope of stellar number density with radius in the central stellar cusp, which can affect relative rates at the factor $\sim 2$ level. 
A factor that could modify this conclusion is the differential mass segregation of stellar sub-populations. In essentially creating differently-sloped radial profiles of different stars (for example white dwarfs versus main-sequence stars) mass segregation might affect relative disruption rates.  Perhaps more significantly, if a substantial population of stellar-mass BHs exists in the cluster, these could possibly or entirely occupy a portion of the tightly-bound phase space from which white dwarf disruptions are thought to arise \citep[see, for example, the analysis of][who study the related occurrence of extreme-mass ratio gravitational wave inspirals]{2016ApJ...830L...1A}. More detailed work is needed to make quantitative statements in the context of white dwarf tidal disruptions.   

The event rate of tidal disruptions by globular cluster IMBHs is thought to be considerably lower than that of galactic nuclei, on the order of $10^{-7}$~yr$^{-1}$ \citep{Ramirez2009}. In this context, N-body stellar dynamics has been the primary tool for analyzing disruption rates. We briefly review these results, with a particular focus on the ways in which they differ from the nuclear-cluster models above. 

The fundamental reasons for the use of N-body dynamics simulations in the globular cluster regime is closely related to the order of magnitude lower disruption rate predicted. The IMBH and its closely bound companion objects have a binding energy comparable to the entire cluster system. Scattering encounters act to keep the cluster core inflated to higher velocity dispersion and lower density \citep[e.g.][]{Baumgardt2004a,Baumgardt2004b,MacLeod2016a}. As a result of this feedback process, the presumption of an homogeneous stellar surrounding, with fixed velocity dispersion $\sigma$, is typically not realized. A related aspect of cluster--IMBH dynamics is that the IMBH wanders relative to the cluster center of mass \citep[for general discussion and a recent analysis based on N-body simulations see, respectively,][]{2013degn.book.....M,2018MNRAS.475.1574D}. A priori, this is not thought to significantly impact white dwarf disruption dynamics, because disrupted white dwarfs tend to arise from orbits that are tightly bound to the IMBH and wander relative to the cluster center with the black hole \citep{MacLeod2016a}. However, more systematic investigation may be warranted, especially if black hole masses as low as $10^3 M_\odot$ are considered. Despite these numerous differences due to overall mass scale between nuclear cluster and globular cluster dynamics we again expect of the order of 1\% of the disruptions that occur will be white dwarf disruptions. The overall white dwarf disruption rate may, therefore, be of the order of 1~Gyr$^{-1}$ per IMBH hosting globular cluster. This order of magnitude event rate has been derived from N-body simulations   \citep{Baumgardt2004a,Baumgardt2004b,MacLeod2016a} and applied to white dwarf disruption scenarios \citep{Haas2012a,Shcherbakov2013a,Sell2015}. Mass segregation effects may have a stronger influence on relative event rates in this context, because globular cluster relaxation times are short relative to the age of the universe. To give a particular example, \citet{Baumgardt2004b} find that white dwarfs make-up approximately 8\% of tidal disruptions, likely because massive white dwarfs segregate into preferentially close orbits to the IMBH when few other massive remants are present.

\subsection{Occurrence of different classes of encounters}
We have, in the preceding section, outlined a variety of classes of white dwarf tidal disruption encounters. White dwarfs have a compactness that implies that gravitational radiation and tides act with similar strengths. Their susceptibility to explosive nuclear burning under degrees of compression that correspond to the BH tidal field further diversifies the potential outcomes. The relative occurrence of these events is a problem that lies at the interface of stellar dynamics and the processes of the events themselves. 

\citet{Zalamea2010} first discussed the partial tidal stripping of white dwarfs over the course of a number of orbits with corresponding gravitational radiation (in the LISA frequency range). \citet{MacLeod2014b} estimated the relative occurrence rate of these events compared to single-passage disruptions to be approximately 10\% of white dwarf disruptions might occur through such a channel. However, the precise proportions remain uncertain (See Fig.~7 of \citet{MacLeod2014b}). 

Deeply-plunging encounters, which can only occur for sufficiently low BH masses, $\lsim 10^5$~\msun, occur when the impact parameter is of the order of $\beta_{\rm thermo} \gsim 3$  \citep[e.g.][]{rosswog2009ab,Anninos2018}. By comparison, white dwarfs begin to lose mass at an impact parameter of $\beta_{\rm ml}\approx 0.5$. The fact that, for most parameters, the phase space of the white dwarf loss cone is thought to be efficiently refilled implies a distribution of impact parameters $N(>\beta)\propto \beta^{-1}$.  In this limit, the fraction of disruptive events that pass deeply enough to lead to a thermonuclear explosion is $\beta_{\rm ml}/\beta_{\rm thermo} \sim 1/6$.

\subsection{Volumetric event rates}
\label{sec_volummetric}

The volumetric rate of white dwarf TDEs is very uncertain -- mostly because the volumetric density of massive BHs below $10^6$~\msun\ is highly uncertain. Converting the specific rate per black-hole cluster system to a volumetric rate involves estimating the volume density of potential black-hole hosts and the BH occupation fraction in those hosts. Explicitly, $\dot N_{\rm vol} \approx \dot N_{\rm BH} n_{\rm BH}$, where $\dot N_{\rm BH}$ is the white dwarf TDE rate per BH (in a given mass range) and $n_{\rm BH}=n_{\rm host}\,f_{\rm BH}$ is the space density of those BHs, which is the product of the density of the hosts and the BH occupation probability for that host type. In view of the importance of the density of massive BHs, we begin the discussion of volumetric event rates by summarizing estimates of the density of potential hosts, namely dwarf galaxies and globular clusters. 

An initial, theoretical estimate of $n_{\rm BH}$ can be made by a flat extrapolation of the massive BH mass function \citep[following][]{MacLeod2016a}. Based on the results of the Illustris simulation at $10^6$~\msun\ \citep[see][]{Sijacki2015} one may infer that $n_{\rm BH}\sim 10^7$~Gpc$^{-3}$. Assuming a per BH rate of $\dot N_{\rm BH}=10^{-6}\;{\rm yr}^{-1}$ (see Section~\ref{sec_dynamics}), the implied volumetric event rate is, therefore, $\dot N_{\rm vol} \approx  10~{\rm yr^{-1}~Gpc^{-3}}$. By comparison, empirical estimates based on the luminosity function or mass function of galaxies give a considerably higher density, if one assumes a high occupation fraction.

The luminosity function of galaxies can be described over the full mass range by a double Schechter function \citep[e.g., equation~(6) of][]{Baldry12}: 
\begin{equation}
\Phi(m)\, dm = \left[\phi_1\, \left(m\over m_*\right)^{\alpha_1} + \phi_2\, \left(m\over m_*\right)^{\alpha_2}\right]
\,e^{-m/m_*}\,{dm\over m_*}\; . 
\label{eq:phigal}
\end{equation}
In the mass range of dwarf galaxies the exponential is approximately unity and equation~(\ref{eq:phigal}) can be approximated as the sum of two power laws. For the purposes of our discussion, we adopt the parameters of this function reported by \citet{wright17}, namely $\log(m_*/\msun)=10.78$, $(\alpha_1,\alpha_2)=(-0.62,-1.50)$, $(\phi_1,\phi_2)=(2.93,0.63)\times 10^{-3}\; {\rm Mpc^{-3}}$ (assuming a Hubble constant of $H_0=70\;{\rm km\;s^{-1}\;Mpc^{-1}}$). The density of dwarf galaxies can then be found by integrating the above function. To determine suitable limits for the integration we turn to \citet{Reines2013} who studied nuclear activity in dwarf galaxies with $\log(m/\msun)=8.5$--9.5 and estimated BH masses in the range  $\log(M_\bullet/\msun)=4.9$--6.5 based on the widths of the broad emission lines. Thus we integrate between $\sim 10^7\;{\rm and}\;10^{9.5}\;\msun$ in order to include BH masses between $\sim 10^4\;{\rm and}\;10^{6.5}\;\msun$ and we obtain a dwarf galaxy density of  $n_{\rm dwarf}\approx 9\times 10^7$ Gpc$^{-3}$. An alternative approach is to integrate the luminosity function of \citet[][expressed in terms of absolute magnitudes in their equation~8]{blanton2005}. Carrying out the integration between absolute $r$ magnitudes of $-13$ and $-18$ using the values of the free parameters for the ``total'' $r$-band luminosity function in table~3 of \citet{blanton2005}, adjusted to $H_0=70\;{\rm km\;s^{-1}\;Mpc^{-1}}$, we get a dwarf galaxy density of the same order as above, $n_{\rm dwarf}\approx 3\times 10^8$ Gpc$^{-3}$.

The occupation fraction of BHs in these systems is unknown, although some observational constraints have been obtained. Searches for AGNs via spectroscopy or variability yield a lower limit of $f_{\rm BH}>\,$a~few~\% \citep{Reines2013, Moran2014, Baldassare2018} while a statistical study of the X-ray properties of nearby, low-mass galaxies yields $f_{\rm BH}>20$\% \citep{Miller2015}. The resulting volumetric rate, based on the latter, higher dwarf galaxy density and assuming a per BH rate of $\dot N_{\rm BH}=10^{-6}\;{\rm yr}^{-1}$ (see Section~\ref{sec_dynamics}) can be expressed as $\dot N_{\rm vol} \approx  300\,f_{\rm BH}~{\rm yr^{-1}~Gpc^{-3}}$ (or $\dot N_{\rm vol} \gsim \,{\rm a~few}\times 10~{\rm yr^{-1}~Gpc^{-3}}$, in view of the limits on $f_{\rm BH}$).

The volume density of globular clusters can be estimated with an analogous approach. A number of recent studies \citep[e.g.,][]{harris13,harris14,zaritsky15} show that the number of globular clusters per massive galaxy ($m>10^{10}\;$\msun) can be expressed in terms of the mass of their host galaxy as $N_{\rm glob}(m) = N_0\,(m/m_0)^\delta$. Combining this relation with the mass function of galaxies in equation~(\ref{eq:phigal}) we get the following density of globular clusters per unit {\it host galaxy} mass
\begin{eqnarray}
  \!\!\!\!
  \Phi_{\rm glob}(m)\, dm 
  \! =\!  N_0 \left(m_*\over m_0\right)^{\!\delta} 
  \left[\phi_1\left(m\over m_*\right)^{\!\alpha_1+\delta}\!\!\!\!
  +\phi_2\left(m\over m_*\right)^{\!\alpha_2+\delta}\right]
  e^{-m/m_*}\, {dm\over m_*}. ~~
  \label{eq:phiglob}
\end{eqnarray}
Adopting $\log N_0=2.924$, $\log(m_0/\msun)=11.2$ and $\delta\approx 1$ \citep{harris13}, and other parameters as in equation~(\ref{eq:phigal}) and integrating\footnote{The integral of the Schechter function can be expressed in terms of incomplete $\Gamma$ functions, $\Gamma(p,a)=\int_a^\infty x^{p-1}e^{-x}\, dx$.} for galaxy masses $m > 10^{10}\;\msun$ we obtain a globular cluster volume density of $n_{\rm glob}(m>10^{10}\,\msun)=9.5\times 10^8\;{\rm Gpc}^{-3}$. An additional contribution may come from globular clusters associated with lower-mass galaxies. In this lower-mass range, $m=10^8$--$10^{10}\msun$, the number of globular clusters per galaxy is described by a flatter power law with an index $\delta=0.365$ and $\log N_0=1.274$, $\log(m_0/\msun)=9.2$ \citep{harris13}. Moreover, equation~(\ref{eq:phiglob}) can be simplified by neglecting the exponential, which is approximately unity for $m < 10^{10}\msun$. Integrating $\Phi_{\rm glob}(m)$ from equation~(\ref{eq:phiglob}) in this range of galaxy masses leads to a density of $n_{\rm glob}(m\! =\! 10^8\textrm{--}10^{10}\,\msun)=4.4\times 10^9\;{\rm Gpc}^{-3}$. But it should be borne in mind that the contribution from lower-mass galaxies is rather uncertain because the small number of globular clusters per low-mass galaxy leads to uncertain parameters for $N_{\rm glob}(m)$ in this mass range.

Searches for IMBHs have put constraints on the BH occupation fraction in globular clusters. These studies can involve measuring the dynamical properties of stars orbiting a putative IMBH in a globular cluster \citep[e.g.,][]{1976ApJ...208L..55N,2003ApJ...589L..25B,2005ApJ...634.1093G,Kiziltan17,Perera17} or measuring accretion signatures from TDEs in globular clusters, such as the IMBH candidate ESO~243-49 HLX-1 or other, similar objects \citep[e.g.,][]{Maccarone2007,2012Sci...337..554W,Lin18}. Other studies have found no signatures of accretion in large samples, suggesting that the typical masses of IMBHs in globular clusters may be lower than theoretical predictions or the occupation fraciton may be extremely small \citep[e.g.][]{Strader2012,2018ApJ...862...16T}. Using a per BH rate of $\dot N_{\rm BH}=10^{-9}\;{\rm yr}^{-1}$ (see Section~\ref{sec_dynamics}) we can express the volumetric rate as $\dot N_{\rm vol} \approx  4\,f_{\rm BH}~{\rm yr^{-1}~Gpc^{-3}}$ or $0.2\,f_{\rm BH}~{\rm yr^{-1}~Gpc^{-3}}$ depending on whether or not we include globular clusters associated with dwarf galaxies. 

 As the above discussion and estimates suggest, the current balance of evidence favors the cores of low-mass galaxies as the most prevalent hosts of white dwarf tidal disruptions. 
It is also worth emphasizing that, since BHs in globular clusters are expected to be less massive than those in dwarf galaxies the detection volume for globular cluster events should be smaller compared to dwarf galaxies because of the lower luminosities of these events.

\subsection{Predicted Detectability in Surveys}
\label{sec_surveys}

In the section, we discuss the probability of the detection of white dwarf TDEs by IMBH using current X-- and $\gamma$--ray missions, as well as in the future next-generation optical transient survey, the Large Synoptic Survey Telescope (LSST).

\subsubsection{Detecting beamed emission with X-- and $\gamma$--ray satellites}
\label{sec_highenergy}

If white dwarf TDEs launch luminous jets, these will be detectable by wide-angle monitors. For example, for a jet luminosity of $10^{48}$~erg~s$^{-1}$, {\it Swift}'s Burst Alert Telescope (BAT) will detect the event out to redshift $z\approx1$, or a luminosity distance of 6.5~Gpc \citep[WMAP 9 Cosmology,][]{MacLeod2016a}. The volume enclosed by redshifts $z<1$ is approximately 150~Gpc$^{3}$. Of the population of events in this volume (see the section above) a fraction is beamed toward our observation perspective, $f_{\rm beam}$, and the BAT detects a fraction of these $f_{\rm BAT}\sim 20$\% because of its sky coverage. The detectable rate is then 
\begin{equation}
\dot N_{\rm BAT}\approx \dot N_{\rm vol} V f_{\rm beam} f_{\rm BAT},
\end{equation}
where $V$ is the observable volume. Due to relativistic abberation, light is emitted in a narrow beam of width 1/$\Gamma^2$, where $\Gamma$ is the Lorentz factor. Fig.~\ref{macleod_2016} assumes $\Gamma\sim7$ implying $f_{\rm beam}\sim1/50$. However, the Lorentz factors associated with relativistic TDEs may be of the order of 2 (\citealt{Zauderer2011a}; \citealt{Cenko2012b}; see the contribution of Zauderer et al.~to this Volume), implying a significantly larger beaming fraction of $f_{\rm beam}\sim 0.1$. Using fiducial numbers and $f_{\rm beam}\sim 1/50$ (and therefore implicitly assuming that all white dwarf TDEs lead to a jet of $10^{48}$~erg~s$^{-1}$) we find $\dot N_{\rm BAT}\approx 6$~yr$^{-1}$ by extrapolation of the BH mass function or $\dot N_{\rm BAT}\approx 180f_{\rm BH}$~yr$^{-1}$ from the dwarf galaxy space density. 

Similarly, sensitive X--ray satellites such as {\it Chandra}, XMM--{\it Newton}, and {\it Swift} might serendipitously detect the X--ray flare thought to accompany a white dwarf TDE. If we again assume that the white dwarf TDE peak  X-ray luminosity L$_X\sim$10$^{48}$erg s$^{-1}$  (see  Fig.~\ref{macleod_2016}), then the volume probed for a conservative flux limit of  10$^{-10}$ erg cm$^{-2}$ s$^{-1}$ is approximately 10$^3$ Gpc$^3$. Again using the volumetric estimates of Section \ref{sec_volummetric}, we estimate
\begin{equation}
\dot N_{\rm X-ray}\approx \dot N_{\rm vol} V f_{\rm beam} f_{\rm sky},
\end{equation}
where $f_{\rm sky}$ is the instantaneous sky fraction observable by the X--ray satellites. This number is of the order of 6$\times 10^{-6}$ for XMM--{\it Newton}'s pn camera with field of view of 30$^\prime \times 30^\prime$. The number is slightly smaller for the {\it Swift} XRT and the {\it Chandra} ACIS--I detector. This fraction also assumes nearly-continuous observing. For $f_{\rm beam}=1/50$, the implied detection rates are  $\dot N_{\rm X-ray} \approx 1.2\times 10^{-3}$~yr$^{-1}$ by extrapolation of the BH mass function or $\dot N_{\rm X-ray} \approx 3.6\times10^{-2} f_{\rm BH}$~yr$^{-1}$ from the dwarf galaxy space density. Thus, for $f_{\rm BH}\sim 1$, we could expect of the order of one detection by  XMM--{\it Newton}, {\it Chandra}, or {\it Swift} over their roughly 20-year missions. 

\subsubsection{Detecting Thermonuclear Transients with LSST}
\label{sec_thermonuc}

Thermonuclear transients produced by the tidal disruption of a white dwarf by IMBH are well within the reach of the planned optical survey of LSST. LSST can detect a prototypical thermonuclear transient, such as that shown in Fig~\ref{macleod_2016}, to $z=0.37$ or a volume of 13.3~Gpc$^{3}$ \citep{MacLeod2016a}. We estimate the detectable rate of white-dwarf TDE thermonuclear transients as 
\begin{equation}
    \dot N_{\rm LSST}\approx \dot N_{\rm vol} V f_{\rm thermo}f_{\rm sky},
\end{equation}
where, $f_{\rm sky}\sim 0.5$, and thermonuclear transients are believed to occur in roughly $f_{\rm thermo}\sim 1/6$ of white dwarf TDEs, see Section \ref{sec_dynamics}. The resultant rates are $\dot N_{\rm LSST} \approx 11$~yr$^{-1}$ based on the extrapolation of the BH mass function, or or $\dot N_{\rm LSST} \approx 330 f_{\rm BH}$~yr$^{-1}$ from the dwarf galaxy space density. There is slight viewing angle dependence introduced to these baseline estimates by the asymmetric ejecta of the tidal thermonuclear transient. See \citet{MacLeod2016a} for a Monte Carlo calculation that shows the preference for certain viewing angles. 

A much greater challenge than detection of tens to hundreds of events per year will be their classification. With LSST anticipated to find hundreds of thousands of supernovae per year, only the multi-wavelength signatures of these white dwarf disruptions make them truly distinctive. One might imagine that the coincidence of high-energy and optical emission (which occurs in a smaller subset of cases, but is more distinctive) might offer one of the only paths to firm classification. 

\section{Multi-Messenger Signatures}
\label{sec_multimsgr}

\subsection{Gravitational Waves from White Dwarf Inspirals Before Disruption}
\label{sec_gw}

Since the disruptions of white dwarfs are often relativistic (i.e., the white dwarf passes within a few Schwarzschild radii from the BH), we can expect substantial emission of gravitational waves.  In general terms, the frequency of gravitational waves emitted during such an encounter is of the same order as the orbital/dynamical frequency at the tidal disruption radius. This time scale is given by equation~(\ref{eq:tpass}) and is comparable to the dynamical time of the star. For the typical white dwarf parameters considered here, $\tau_{\rm dyn}\approx 3.5\; (\Mwd/0.6\;\msun)^{-1/2}\; (\Rwd/10^9\;{\rm cm})^{3/2}\;$s, implying gravitational wave frequencies on the order of $f_{\rm gw}\sim 2/\tau_{\rm dyn} \sim 0.6\;$Hz. This frequency is of the same order as the range accessible by the planned LASER Interferometer Space Antenna \citep[hereafter LISA; anticipated launch date in the mid 2030s, see][]{Danzmann17}.

In the more likely case of a white dwarf in an unbound, parabolic orbit a burst of gravitational waves will be emitted during the initial encounter with the BH. After the initial encounter the debris will return to pericentre one dynamical time later and accretion will begin soon thereafter, leading to an electromagnetic flare. The gravitational wave strain (effectively, the observed amplitude of the wave or pulse) was computed by \citet[][see their \S{6} for the details of the calculation and relevant formulae and references]{rosswog2009ab} who found it to be on the order of $h\sim10^{-19}$, for a 0.6~\msun\ white dwarf passing at the tidal disruption radius of a $10^3\;$\msun\ BH at a distance of 10~kpc.  This combination of parameters is appropriate for a Milky Way globular cluster, and suggests that the burst of gravitational waves from the disruption of a white dwarf in a nearby globular cluster may be detectable. However, more distant events, outside of the local group, are probably not detectable \citep[see discussion in][]{Anninos2018}.

In an alternative and possibly rarer scenario the white dwarf is captured in a bound, typically eccentric, orbit around a BH and then spirals in. The per BH rate of such events is 1/10 the rate of all white dwarf disruptions (see discussion in Section~\ref{sec_volummetric}). Such events are also referred to as ``extreme mass ratio inspirals'' (EMRIs, BH to white dwarf mass ratio $>10^4$) or  ``intermediate mass ratio inspirals'' (IMRIs, BH to white dwarf mass ratio $10^2$--$10^4$). They are interesting because the white dwarf acts as a test particle that executes a very large number of orbits in the potential of the BH, hence the gravitational wave signal allows one to ``map'' the  space-time around the BH and determine its fundamental properties: its mass and spin \citep[see, for example, the review by][]{amaro07}. 

Since the white dwarf will complete a large number of orbital cycles around the BH before it is disrupted ($10^5$ revolutions, or even more) the gravitational wave signal is continuous and, in fact, the amplitude of the wave increases along with its frequency, as the white dwarf spirals in. The gravitational wave strain for an IMBH in a nearby (50--100~Mpc) dwarf galaxy is $h\sim10^{-22}$--$10^{-21}$. The slow inspiral allows for a detection of the gravitational wave signal with a high signal-to-noise ratio (hereafter $S/N$) by accumulating a large number of orbital cycles in a long ``exposure'' or ``integration.'' \citet{Sesana2008} have estimated \citep[using the methodology described in][]{Barack2004} the $S/N$ attainable by a detector with capabilities similar to those of LISA and found that such systems may be detectable with $S/N > 30$ at a distance of 100~Mpc a few years before the disruption (with an integration time of order years). Under favourable conditions (i.e., a final orbital eccentricity of zero and an integration time of 3--5 years) detections of more distant systems, up to 450~Mpc away ($z\lsim 0.1$), are possible, although with a lower $S/N$. Since the orbit is (at least initially) eccentric, the higher harmonics of the signal will be stronger than the fundamental and their detection will allow determination of the fundamental system parameters, i.e., the masses of the objects involved, the distance, and the BH spin. Knowing these parameters is extremely useful for using TDEs as tools to study BH properties (as discussed in Section \ref{sec_intro}).

A typical event involving the inspiral of 0.6~\msun\ white dwarf into a 1$0^5\;$\msun\ IMBH begins with the capture of the white dwarf in a highly eccentric, bound orbit with a pericentre distance of $\sim 15\;\Rg$ ($\sim 2\times 10^{11}\;$cm), corresponding to an orbital period of 3 minutes and an initial gravitational wave frequency of $f^{\rm init}_{\rm gw}\approx 10\;$mHz \citep[see, for example,][]{Sesana2008,MacLeod2014b,MacLeod2016b}. The orbit decays gradually via the emission of gravitational waves with a possible contribution from tidal heating and radiative cooling of the white dwarf \citep[see][]{MacLeod2016a}. The tidal heating can  be amplified by resonances \citep[e.g.,][]{Rathore05} and can lead to substantial brightening of the white dwarf and may trigger runaway fusion in a surface layer of hydrogen \citep{Vick2017}, possibly producing a soft X-ray flare. If emission of gravitational radiation is the dominant orbital decay mechanism, the decay time is approximately 3.4~years. During this period, the system is a steady source of gravitational waves whose frequency evolves to a maximum value of $f^{\rm fin}_{\rm gw}\approx 120\;$mHz, as the orbital separation reaches the tidal disruption radius of $\Rt\approx 3\;\Rg$. Just before the white dwarf is disrupted, it fills its Roche lobe and goes through a phase of (stable) mass transfer onto the BH that can last days to weeks \citep[see][]{Zalamea2010,Dai2013b}. The accretion of this gas by the BH can lead to a substantial luminosity, on the order of $10^{43}\;{\rm erg\; s}^{-1}$, presumably in the X-ray band. If the orbit is still eccentric at this stage, the mass transfer, hence the X-ray emission, will also be periodic. The above scenario leads to intriguing predictions for the observational signature of this event. The first signal to be detected is the gravitational wave signal, which allows us to infer the masses of the objects involved, although the location in the sky will be fairly uncertain. This signal will also give us early warning of the TDE. The tidal heating and brightening of the white dwarf may lead to an observable, periodic signal in the optical/UV band (this is especially likely if nuclear burning is triggered on the surface of the white dwarf). Then, a few days to weeks before the ultimate disruption, we may detect steady or periodic X-ray emission as the white dwarf loses mass via Roche lobe overflow. The combination of these observational signatures holds great diagnostic power.

\subsection{Ultra-High Energy Cosmic Rays from TDE Jets and Other Outflows}
\label{sec_cr}
The jets and other outflows produced in some TDEs are also sites of particle acceleration and have been suggested as the sources of ultra-high energy cosmic rays (UHECRs, $E \gsim 10^{18}\;$eV) and neutrinos \citep[see, for example][]{Farrar09,Farrar2014a,Wang16,Senno17,Lunardini17}. Here we focus on the special conditions available in white dwarf TDEs and how they affect the process of UHECR production. The special feature of white dwarf TDEs is that they release debris that is rich in elements heavier than helium (hereafter, ``metals''), which means that the resulting jets are made up of metals. Therefore, an appreciable number of metal nuclei can survive shocks internal and external to the jet and eventually arrive at the Earth and be detected by ground-based detectors as UHECRs. This particular feature of white dwarf TDE jets is essential in explaining the composition of UHECRs.

Three recent studies have computed the composition and energy spectrum of particles that are accelerated in white dwarf TDE jets and detected as UHECRs and neutrinos: \citet{Zhang17}, \citet{Alves17}, and \citet{Biehl18}.  These authors considered the acceleration of the particles, the reactions that may destroy metal nuclei, the propagation of the particles from the source to the Earth and their detection by the IceCube and Pierre Auger observatories. The general conclusion is that the models can reproduce both the composition and energy spectrum of UHECRs \citep[][note that O-Ne-Mg white dwarfs lead to better agreement]{Zhang17}. The implied TDE rates range between 0.1~yr$^{-1}$~Gpc$^{-1}$ \citep{Biehl18} and 10~yr$^{-1}$~Gpc$^{-1}$ \citep{Alves17} but these are inversely proportional to the baryon loading  of the jets (the ratio of proton to photon luminosities) and proportional to the IMBH occupation fraction (see Section~\ref{sec_volummetric}). In view of the uncertainties in these parameters, the rates reported in this paragraph are not incompatible with those of Section~\ref{sec_volummetric}.

\section{Conclusions}
This chapter provides an overview of the tidal disruption of white dwarfs by IMBHs. While no unambiguous white dwarf TDE has yet been identified, theoretical calculations and simulations suggest that if IMBHs exist, then white dwarfs will be tidally disrupted by them if they approach close enough. The subsequent detonation of the white dwarf by tidal compression is dependent on the conditions of the system but some are expected to produce observable signatures at optical wavelengths, as well as high-energy counterparts. A number of candidate events have been suggested that match many of the predicted properties but no clear case as yet been observed. The predicted intrinsic rates of white dwarf TDEs are relatively uncertain but current and future transient surveys, such as LSST, should discover them at levels where identification is possible. This will allow the physics of these extreme systems to be studied along with the implications for populations of IMBH formation and evolution. 


\bibliographystyle{aps-nameyear}      
\bibliography{general.bib}               
\end{document}